\documentclass[preprint2]{aastex63}
\shortauthors{Huber et al.}
\graphicspath{{./}{figures/}}
\usepackage{float}
\usepackage{longtable}
\usepackage{lineno}
\DeclareUnicodeCharacter{2212}{-}
\begin{document}
\title{The Hosts of X-Ray Absorption Lines Toward AGNs}
\author[0000-0001-7829-4764]{Maggie C. Huber}
\affiliation{Department of Astronomy, University of Michigan \\
Ann Arbor, MI 48109, USA}
\email{maghuber@umich.edu, jbregman@umich.edu}

\author[0000-0001-6276-9526]{Joel N. Bregman}
\affiliation{Department of Astronomy, University of Michigan \\
Ann Arbor, MI 48109, USA}
\begin{abstract}
Most baryonic matter in the universe exists in gaseous form and can be found in structures such as galactic halos and the low-density intergalactic medium. Proposed X-ray spectroscopy missions such as \textit{Athena}, \textit{Arcus}, and \textit{Lynx} will have the capability to identify absorption lines in spectra toward bright active galactic nuclei (AGNs), which can be used as a tool to probe this missing matter. In this study, we examine the optical fields surrounding 15 primary observing targets and identify the foreground galaxies and galaxy groups that are potential hosts of absorption. We record the basic properties of the potential host and their angular and physical separation from the AGN line of sight. This process is done by marking the location of various galaxies and groups in optical images of the field surrounding the target and plotting their angular separation vs. redshift to gauge physical proximity to the background source. We identify the surrounding objects according to those which have measured redshifts and those that require them. 
\end{abstract}


\section{Introduction} \label{sec:intro}
In spectra of distant quasars, gaseous material from galaxies along the line of sight causes absorption lines which provide a precise way to examine the baryonic matter of galaxies. Quasar absorption line studies have revealed that most baryons in the Universe are in gaseous form and in a variety of structures, from galaxy disks and halos to the lower density intergalactic medium (IGM) \citep{1998ApJ...503..518F}. These studies are carried out in the ultraviolet, optical, and radio bands, which are sensitive to gas in the $10$–
$10^{5.5}$ K range. The baryons accounted for in these studies result in an overall census that falls 30-50\% short of what is predicted by Big Bang Nucleosynthesis \citep{2012ApJ...759...23S}. It is important to note the temperature range of these studies are below the virial temperature of L* galaxies ($\approx 10^{6.3}$ K), or the common galaxy groups ($10^{6.5}$-$10^{7.5}$ K).  In these higher temperature ranges, the prominent resonance absorption lines lie in the X-ray band (0.3-0.8 keV) \citep{2011ApJ...731....6S}. Observations of the Sunyaev-Zeldovich effect in locally brightest galaxies \citep{2013A&A...557A..52P,breg2021} imply that we should observe most missing baryons at the virial temperatures of galaxies. Thus, we expect to account for the missing baryons by extending absorption line studies to the X-ray waveband where we can probe these temperatures. This contradicts other explanations based on UV absorption lines that place the majority of baryons in the circumgalactic medium (CGM) at temperatures of 10$^{4}$ K \citep{2014ApJ...792....8W}. The results of X-ray spectroscopy should be able to clarify to some extent whether this is the case or if virial temperatures of galaxies dominate the distribution of baryons. 

X-ray absorption line studies of this hot halo material in galaxies and galaxy groups are in their infancy due to the modest spectroscopic capabilities of current X-ray observatories, such as \textit{XMM-Newton} and the \textit{Chandra X-ray Observatory} \citep{2012ApJ...746..166Y}.  Milky Way halo gas is detected by these observatories in several X-ray absorption lines, from which one obtains information about the extent of the halo and thick disk, its composition, and that the extended halo rotates \citep{2013ApJ...770..118M}.  Only one extragalactic absorption line was detected \citep{nicastro18}, and a likely galaxy host has been identified. X-ray and UV spectroscopy have also been successful in revealing the physical origins of the warm-hot intergalactic medium \citep{2019ApJ...884L..31J} and intragroup gas \citep{2002ApJ...580..169B} through absorption lines in the spectra toward AGNs.

Although current X-ray observatories do not possess the spectral resolution or collecting area to probe the virial temperatures of L* galaxies, we should be able to achieve this soon. Greatly improved X-ray absorption line sensitivity will be possible with upcoming missions, such as \textit{Athena}, \textit{Arcus}, and \textit{Lynx} \citep{barcons17,2019JATIS...5b1001G,2020SPIE11444E..2CS}.  These observatories will likely observe more than $100$ extragalactic absorption line systems in their first few years of operation.  The objects that will provide the best continuum to measure absorption against are AGNs, especially blazars. The highest priority background sources are already identified, based on mean X-ray brightness and redshift; nearly all are blazars \citep{2015JATIS...1d5003B}. In most of these blazar fields, the potential galaxy and galaxy group hosts have yet to be identified. It is important to note that progress has been made by dedicated efforts to conduct surveys of galaxies surrounding UV bright quasars \citep{2011ApJS..193...28P, 2019ApJS..243...24P, 2019ApJ...884L..31J}.

Our goal is to identify likely hosts along the lines of sight toward the blazars most likely to be observed with these new observatories.  Identification is helped considerably by the various optical and infrared surveys that have been carried out, including the \textit{Two Micron All-Sky Survey} \citep{2MASS2006}, the \textit{Wide-field Infrared Survey Explorer} \citep{WISE2010}, the \textit{Sloan Digital Sky Survey} \citep{SDSS2017}, \textit{Pan-STARRS} \citep{PanStarrs2020}, the \textit{Dark Energy Survey} \citep{DES2021} and the \textit{VLT-ATLAS} survey \citep{VLT_ATLAS2015}.

Here, we examine each high priority field and identify possible hosts that already have redshifts and others that need redshifts. Section 2 covers the methods and selection criteria, section 3 goes through each field individually, section 4 discusses applications of this study and future work, and finally section 5 summarizes our findings.
\newline

\section{Methods} \label{sec:methods}
In Table \ref{table1}, we show the observing targets that we cover in this study chosen from a previously curated list of 104 AGN sources \citep{2015JATIS...1d5003B} that provide the brightest X-ray continuum for detecting absorption lines. 

\begin{deluxetable*}{lllll}
\tablenum{1}
\tablecaption{Summary of the AGN sources we selected for this study.}
\tablewidth{0pt}
\tablehead{ \colhead{Object Name}  & \colhead{Right Ascension}  & \colhead{Declination}  & \colhead{Redshift} & \colhead{Merit}
}
\startdata
1RXS J003334.6-192130 & 00:33:34.200  & -19:21:33.30  & 0.610 & 4.08E-12 \\ 
1RXS J022716.6+020154 & 02:27:16.580  & +02:02:00.50  & 0.457 & 5.92E-12 \\ 
S5 0836+71            & 08:41:24.3652 & +70:53:42.173 & 2.172 & 6.70E-12 \\ 
1ES 1028+511          & 10:31:18.518  & +50:53:35.82  & 0.360 & 5.94E-12 \\ 
1RXS J110337.7-232931 & 11:03:37.610  & -23:29:31.20  & 0.186 & 5.32E-12 \\ 
Mrk 421               & 11:04:27.3139 & +38:12:31.799 & 0.030 & 7.44E-12 \\ 
PG 1553+113           & 11:04:27.3139 & +11:11:24.365 & 0.450 & 8.32E-12 \\ 
1RXS J111706.3+201410 & 11:17:06.260  & +20:14:07.40  & 0.139 & 4.05E-12 \\ 
1RXS J122121.7+301041 & 12:21:21.941  & +30:10:37.11  & 0.184 & 4.19E-12 \\ 
3C 273                & 12:29:06.6997 & +02:03:08.598 & 0.158 & 8.20E-12 \\ 
Ton 116               & 12:43:12.7362 & +36:27:43.999 & 1.065 & 8.97E-12 \\ 
3C 279                & 12:56:11.1665 & -05:47:21.523 & 0.536 & 4.14E-12 \\ 
1RXS J142239.1+580159 & 14:22:38.895  & +58:01:55.50  & 0.635 & 6.48E-12 \\ 
1RXS J150759.8+041511 & 15:07:59.7324 & +04:15:11.984 & 1.703 & 4.80E-12 \\ 
1RXS J151747.3+652522 & 15:17:47.600  & +65:25:23.90  & 0.702 & 6.34E-12 \\ 
1RXS J153501.1+532042 & 15:35:00.801  & +53:20:37.33  & 0.890 & 8.52E-12 \\ 
PKS 2155-304          & 21:58:52.0651 & -30:13:32.118 & 0.116 & 2.14E-11 \\ 
3C 454.3              & 22:53:57.7479 & +16:08:53.560 & 0.859 & 4.93E-12 \\ 
H2356-309             & 23:59:07.910  & -30:37:40.60  & 0.165 & 3.83E-12 
\label{table1}
\enddata
\tablecomments{Merit taken from \citep{2015JATIS...1d5003B}. Targets 3C 273, Mrk 421, PG 1553+113, and PKS 2155-304 will not be discussed in the results section this paper as the surrounding fields have already been studied in detail. We mention them in this table because they are included in the primary targets of future X-ray observatories. Information about these fields can be found at (in order): \cite{2011ApJS..193...28P}, \cite{2011ApJ...736..131A}, \cite{2019ApJ...884L..31J},\cite{2019MNRAS.484..749C}.}
\end{deluxetable*}

We gather information about foreground objects by conducting searches in the \textit{NASA/IPAC Extragalactic Database} (NED) for all objects within a radius of 10 arcminutes with the AGN at the center. The cosmological parameters are set according to the Planck 2015 results \citep{2013A&A...557A..52P} with redshift corrected to the reference frame defined by the 3K CMB. We create tables of NED values for the 
redshift, location (RA and DEC), apparent magnitude and angular separation of each object. The completeness of the NED data varies with each AGN field, and we do not include magnitudes in the results for fields which have incomplete photometry. There are multiple objects in every field that do not have 
available redshifts in NED, so we discuss those objects separately. We also correct our lists for objects that are double counted or false positives (i.e. listing a source at a location where there was not one 
visible in the optical image). Galaxies and stars are differentiated by NED based on their instrumental point spread function (PSF). In the survey data we obtained, galaxies are distinguished by an extended PSF, whereas stars are point-like. We also confirm NED object classifications by inspection of fields and objects with a combination of SAOImage DS9 \citep{DS9_2003J}, ALADIN \citep{ALADIN2000}, and the SDSS Navigation Tool. 

When we have information from NED, we use the \textit{ESO Online Digitized Sky Survey} (DSS) to obtain optical images of these regions from the \textit{Second Generation DSS} in the red waveband. This survey uses the \textit{Oschin} (Palomar Schmidt) and \textit{UK Schmidt} telescopes with a limiting magnitude of approximately 22.5 mag (B$_{J}$) \citep{1991PASP..103..661R}. Using a limit 1.5 mag lower than this value, we filter the list to exclude all objects with a magnitude greater than 21 mag (B$_{J}$) for fields that had over 100 objects in NED. We set the limit to be lower than that of the \textit{Oschin} and \textit{UK 
Schmidt} telescopes so our primary sources are accounted for in other optical surveys with lower limits (i.e., \textit{DES}, \textit{SDSS}, \textit{Pan-STARRS}, \textit{VLT-ATLAS}). The image size is 20$^\prime$x20$^\prime$ by default, with some using different angular sizes to better show the 
distribution of objects in the field. We indicate when the image size deviates from the default in the figure captions. To complete these images, we make a region file for each field, marking the location of 
objects of interest, color coded by redshift if applicable. We superimpose these regions on the optical DSS images using \textit{SAOImage DS9}, and we use these figures as a visual guide to select potential absorbers. We mark the most likely absorbers (criteria for this listed below) with a number and discuss the numbered objects in the results. We use standard cardinal directions with the AGN at the center to refer to the location of an object. 

We also create plots of angular separation in arcminutes as a function of redshift, marking the redshift of the AGN as a vertical line. This is only done for objects that have available redshift values. To make an estimate of the physical distance to the AGN, we use \textit{Cosmology Calculator} \citep{2006PASP..118.1711W} with flat universe parameters to find separation as a function of redshift for physical distances of 1 and 0.5 Mpc. We plot these functions on the graph to gauge an object’s physical distance to the AGN. For example, if the object is closer than 0.5 Mpc to the AGN it lies below the line corresponding to that distance. 

The NED data, optical images, and separation vs. redshift graphs allow us to select the most likely objects to cause absorption lines. The main criteria we look for are objects that are brightest (lowest magnitude) and have the closest proximity to the AGN.
The objects we selected to discuss in the brief paragraphs for each field minimize both the magnitude and separation. We calculate the impact parameter for each object by dividing the physical diameter by the angular diameter then multipyling by the separation. The resulting list of objects provides a good foundation upon which we expand to include other objects of interest. Using the separation vs. redshift graphs, we identify and pick objects that are possible virialized galaxy groups or lie at the redshift of the AGN.

\section{Potential absorption hosts}

Not all galaxies are expected to have extended hot halos that would produce X-ray absorption lines, such as O VII resonance lines from K-alpha emission, which should be the strongest.  Halo mass calculations for an O VII column density 
$ \geq 10^{15}$ cm$^{−2}$
peak at a stellar mass of $M_{star} \approx 10^{10.9}$ M$_{\odot}$, decreasing to higher and lower stellar masses \citep{Qu2018b}.  
In the following, we choose to identify the galaxies more massive than M(r) = -20, which is about 1.5 magnitudes below L* and corresponds to $M_{star} \approx 10^{10.4}$ M$_{\odot}$.
We can convert this into a magnitude as a function of redshift, including a K correction for a typical late-type galaxy, which corresponds to r magnitudes of: 18.4 at z = 0.1; 20.2 at z = 0.2; 21.2 at z = 0.3; 22.0 at z = 0.4; and 22.7 at z = 0.5.  
When z $< 0.2$, SDSS provides photo-z values that are generally accurate and a number of spectra as well.  For z $\approx 0.3$, photo-z values become less reliable, while spectra are lacking for all but the most luminous galaxies.  
However, galaxies can be identified to r $\approx 22$ with SDSS and fainter with Pan-STARRS and DES (magnitude limit of 23.2 - 23.5), so potential M(r) = -20 target galaxies can be identified to z $\approx 0.4$, beyond which dedicated imaging programs would be required. 

We also can apply a characteristic search radius and use 1 Mpc, which is $\sim 5$ R$_{vir}$ of an L* galaxy and about 2R$_{500}$ for a galaxy group or poor cluster.  At a fixed metric radius, the angular search radii decreases until $z \sim 0.5$, after which the angular radius changes slowly.  The implication is that at lower redshift, such as z = 0.1, the search radius is 9$^\prime$ and with m(r) $<$ 18.38 -- the target galaxies are relatively bright and are easy to identify.  The 1 Mpc radius corresponds to 5$^{\prime}$ at z = 0.2, 3.7$^{\prime}$ at z = 0.3, 3.1$^{\prime}$ at z = 0.4, 2.7$^{\prime}$ at z = 0.5, and 2.0$^{\prime}$ at z = 1. This leads to a decreasing search area with increasing z, but with fainter galaxy candidates.  This aspect is included in our analysis of the images. 

\subsection{1ES 1028+511}
\subsubsection{Objects with known redshifts}
\begin{figure}
\centering
\includegraphics[width=0.8\columnwidth]{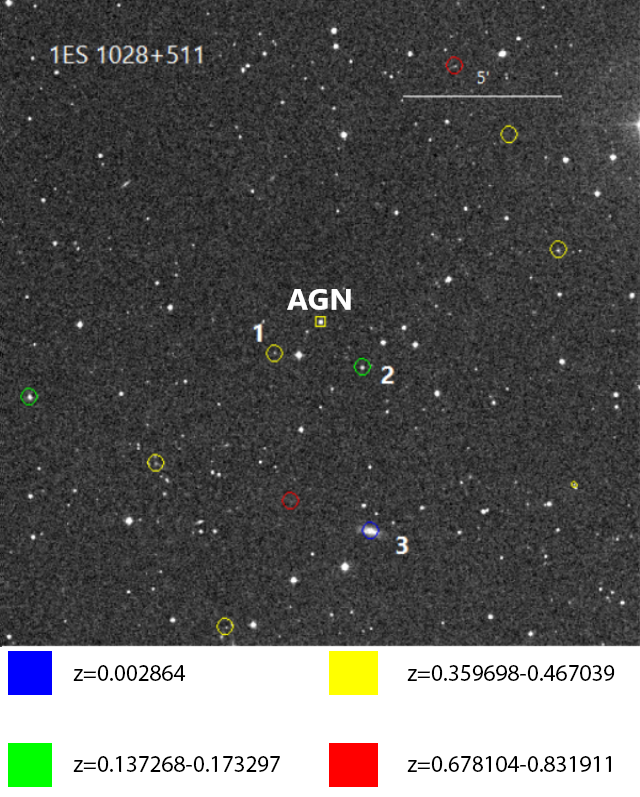}

(a) Figure 1a
\includegraphics[width=0.8\columnwidth]{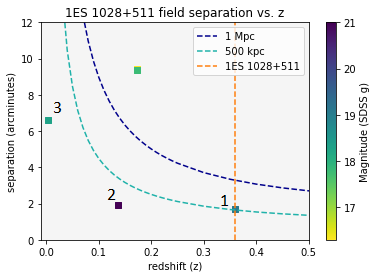}

(b) Figure 1b
\caption{ Cardinal directions in the figure are standard: N is up, S is down and E is to the left in the direction of increasing Right Ascension. (a) The box at the center marks the location of the AGN. Objects within a radius of 10$^\prime$ from the AGN are marked by a circle with color corresponding to their redshift. (b) Separation of the object from the AGN in arcminutes versus the redshift. Dashed lines show 1 Mpc, 500 kpc. 500 kpc is about 2$R_{200}$ of a L* galaxy while 1 Mpc is similar to $R_{200}$ of a medium mass galaxy cluster. A vertical line indicates the redshift of the AGN.}
\label{fig1}
\end{figure}
For the objects with known redshifts surrounding 1ES 1028+511 (z=0.36), we found four candidates for absorption. 2MASS J10312764+5052394 (Object 1: 10:31:27.6,+50:52:39) is the closest object, located 1.7$^\prime$ SE  from the AGN in Figure \ref{fig1} with a spectroscopic redshift of 0.36 and magnitude 21.0 (g SDSS Model AB). It's estimated physical distance is 0.5 Mpc with a line-of-sight impact parameter of 795 kpc. 2MASS J10311034+5052107 (Object 2: 10:31:10.3,+50:52:11) is the next closest object with a spectroscopic redshift of 0.14, magnitude 18.3 and separation 1.9$^\prime$ SW. It has a physical distance of less than 0.5 Mpc and an impact parameter of 339 kpc. An additional prominent object in the field  is SDSS J103108.88+504708.7 (Object 3: 10:31:08.9,+50:47:09) has a spectroscopic redshift of 0.0029, magnitude 16.3 and separation 6.6$^\prime$ SW. This object has a physical distance of less than 0.5 Mpc with an impact parameter of 37.5 kpc. 
\subsubsection{Objects with unknown redshifts}
\begin{figure}[H]%
\centering
\includegraphics[width= 0.8\columnwidth]{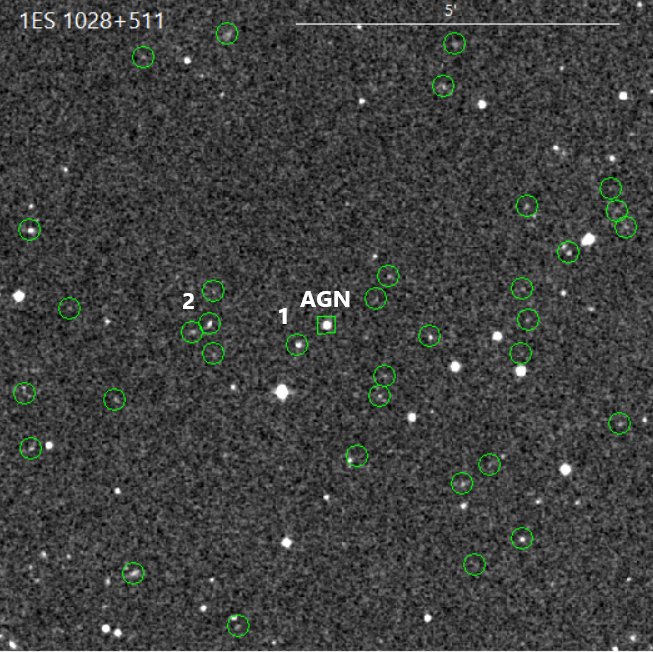}

\caption{   {Potential absorption line host galaxies without known redshifts. The angular size of the image (10$^\prime$) was narrowed from the 20$^\prime$ (angular size of Figure 2a) in order to narrow the number surrounding objects from 197 to 35. }}
\label{fig2}
\end{figure}
The objects with unknown redshifts surrounding 1ES 1028+511 were selected between magnitude 19.0 and 22.0 (SDSS g) to accommodate the magnitude limits of observational telescopes, taking in all absorbers besides objects below the flux threshold of the survey at the redshift of the AGN (Magnitude = -19.0). There is one object and another grouping of 4 objects shown in Figure \ref{fig2} within 2.0$^\prime$ of the AGN that are likely candidates. The object WISEA J103121.30+505317.8 (Object 1: 10:31:21.31,+50:53:17.84) is 0.53$^\prime$ E is closest to the AGN with magntiude 20.0 (SDSS g). 

There are also four objects (Group 2: WISEA J103129.44+505310.5, WISEA
J103129.87+505337.3, WISEA J103129.38+505406.9, 

WISEA J103131.50+505330.2)
located 1.8$^\prime$ E of the AGN around RA 10:31:29.5 and DEC 50:53:37.8 with magnitudes ranging from 19.9 to 20.7. There is a possibility that these four objects are part of a virialized galaxy group. Until redshifts are obtained for the objects in this field we will not be able to determine the physical distance between the objects and the AGN. Information on all objects in this field can be found in Table \ref{tbl3}.

\subsection{1RXS J003334.6-192130}
\subsubsection{Objects with unknown redshifts}
\begin{figure}[H]%
\centering
\includegraphics[ width=0.8\columnwidth]{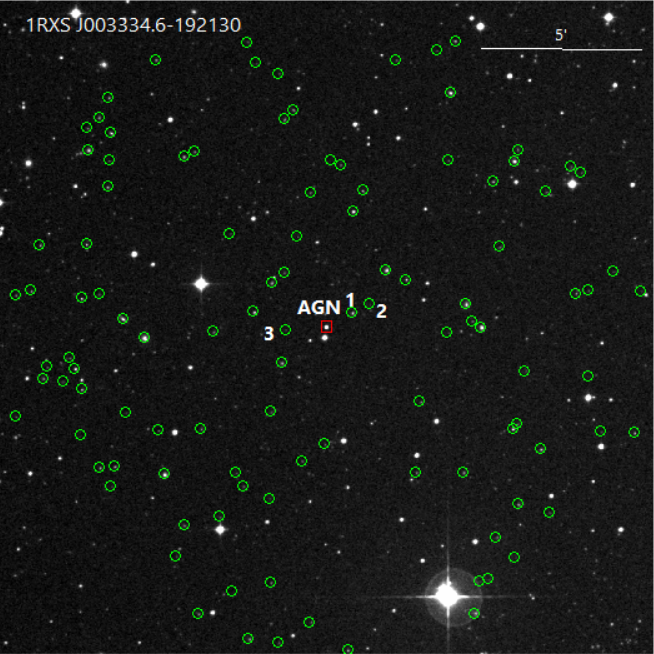}
\label{fig3}
\caption{   }
\end{figure}
This field lies outside the SDSS footprint, resulting zero objects with known photo-z redshift values within 10$^\prime$ of 1RXS J003334.6-192130. There are 108 objects found with unknown redshift and optical magnitude. One source is WISEA J003330.93-192108.8 (Object 1: 00:33:30.94,	-19:21:9.04) located 0.87$^\prime$ NW of the AGN in Figure \ref{fig3}. WISEA J003328.74-192053.4 (Object 2: 00:33:28.73,-19:20:53.92) located 1.5$^\prime$ NW of the AGN. The third object of interest is WISEA J003339.61-192138.0 (Object 3: 00:33:39.65,-19:21:37.91) with separation 1.3 E. Further observations of this field will be necessary to locate more galaxies and finding observational redshifts to obtain a physical distance from the AGN. Information on all objects in this field can be found in Table \ref{tbl4}.

\subsection{1RXS J022716.6+0201541}
\subsubsection{Objects with known redshifts}
\begin{figure}[H]%
\centering
\includegraphics[width=0.8\columnwidth]{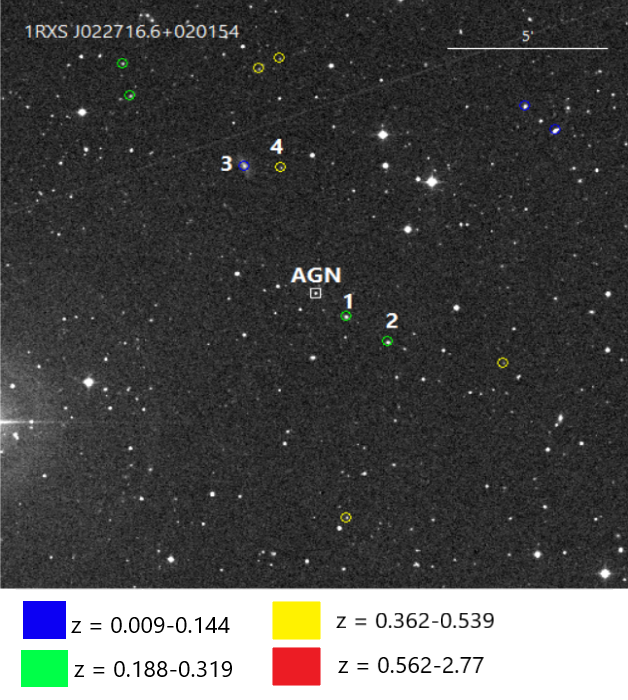} 

(a) Figure 4a
\includegraphics[width=0.8\columnwidth]{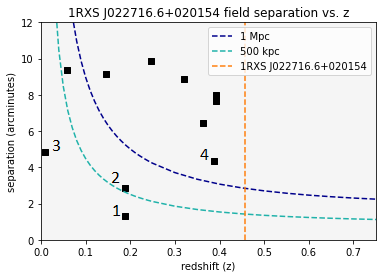}

(b) Figure 4b
\caption{   (b) There is incomplete optical photometry for these sources, so magnitudes are not plotted in Figure 5.2.}
\label{fig4}
\end{figure}
Within a 10$^\prime$ radius of the AGN, there are 12 objects selected that are candidates for absorption in Figure \ref{fig4}. The physically closest objects to the AGN include WISEA J022712.68+020111.6 (Object 1: 02:27:12.62, 02:01:11.28), at spectroscopic redshift 0.19, separation 1.2$^\prime$ SW, and  physical distance of less than 0.5 Mpc. WISEA J022707.38+020019.7 (Object 2: 02:27:7.35, 02:00:19.69) is at spectroscopic redshift 0.19, separation 2.9$^\prime$ SW and physical distance of approximately 0.5 Mpc. These two objects may belong to a galaxy group. UGC 01923 (Object 3: 02:27:25.55, 02:06:18.11) is another ideal candidate at spectroscopic redshift 0.0095 and 4.8$^\prime$ (less than 0.5 Mpc for the redshift of 1RXS J022716.6+0201541, which is z=0.457) NE. The impact parameter for this object is 57.4 kpc. WISEA J022720.93+020614.6 (Object 4: 02:27:20.94,+02:06:14.51) is located 4.4$^\prime$ NE at a spectroscopic redshift of 0.39. It has a physical distance greater than 1 Mpc.

\subsubsection{Objects with unknown redshifts}
\begin{figure}[H]%
\centering
\includegraphics[ width=0.8\columnwidth]{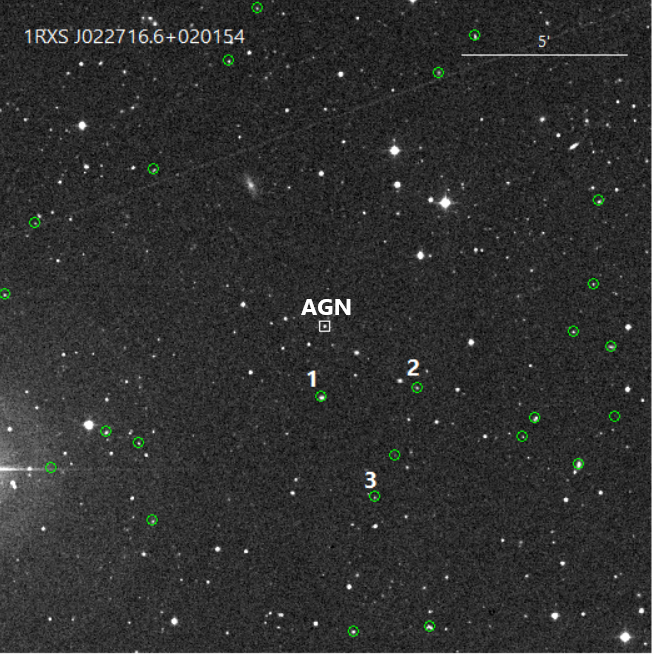}
\caption{   }
\label{fig5}
\end{figure}
There are 25 objects selected within a 10$^\prime$ range of the AGN with unknown redshifts, between magnitude 18.1 and 20.5. The object with the closest angular distance to the AGN is WISEA J022716.94+015948.9 (Object 1: 02:27:16.9,+01:59:49) with magnitude 18.5 (J) located 2.2$^\prime$ S of the AGN in Figure \ref{fig5}. The next closest object is WISEA J022705.29+020006.7 (Object 2: 02:27:05.2,+02:00:06) at magnitude 20.4 (J) and 3.4$^\prime$ SW of the AGN. Another possible candidate for absorption is WISEA J022710.50+015645.5 (Object 3: 02:27:10.45,+01:56:45.46) at magnitude 20.4 located 5.5$^\prime$ SE. Information on all objects in this field can be found in Table \ref{tbl5}. 
%
\subsection{1RXS J110337.7-232931}
\subsubsection{Objects with known redshifts}
\begin{figure}[H]%
\centering
\includegraphics[width=0.8\columnwidth]{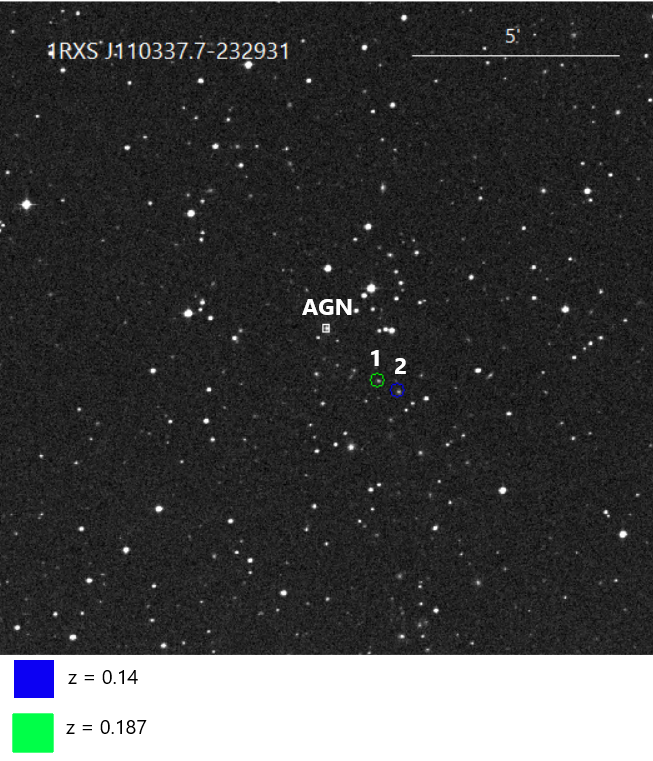} 

(a) Figure 6a
\includegraphics[width=0.8\columnwidth]{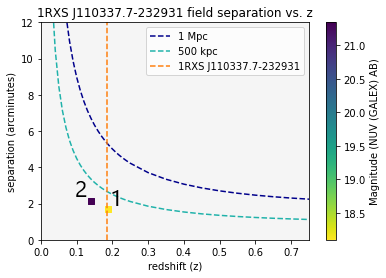} 

(b) Figure 6b
\caption{   }
\label{fig6}
\end{figure}
Like 1RXS J003334.6-192130, this field lies outside the SDSS footprint. Two objects were found within 10$^\prime$ of the AGN which are located 2.0$^\prime$ SW in Figure \ref{fig6}, H 1101-232G1 (Object 1: 11:03:32.5,-23:30:44) with spectroscopic redshift 0.187 with physical distance less than 0.5 Mpc. GALEXASC J110330.49-233059.6 (Object 2: 11:03:30.5,-23:30:58) has a spectroscopic redshift 0.14 and corresponding physical distance of less than 0.5 Mpc.
\vspace{-0.3cm}
\subsubsection{Objects with unknown redshifts}
\begin{figure}[H]%
\centering
\includegraphics[width= 0.8\columnwidth]{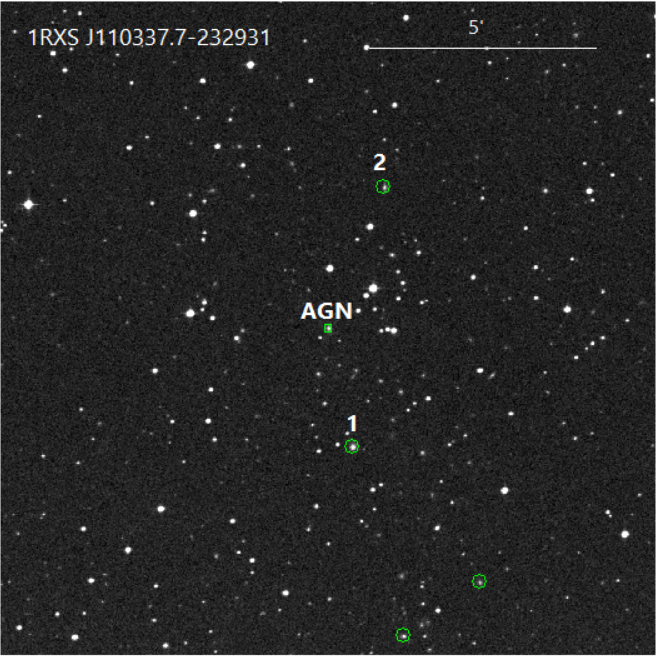}
\caption{   }
\label{fig7}
\end{figure}
There are four objects that were found within 10$^\prime$ of the AGN, and all lack magnitudes. The object closest in angular separation WISEA J110335.26-233214.2 (Object 1: 11:03:35.2,-23:32:14) is located 2.8$^\prime$ S of the AGN in Figure \ref{fig7}. WISEA J110331.98-232617.9 (Object 2: 11:03:32.0,-23:26:18) is also close to the AGN, located 3.5$^\prime$ NW. Due to the poor survey coverage of this region, we may need further observations so that a physical distance from the AGN can be determined. We need followup photometry if this object is a potential target for X-ray spectroscopy missions. Information on all objects in this field can be found in Table \ref{tbl6}.
\subsection{1RXS J111706.3+201410}
\subsubsection{Objects with known redshifts}
\begin{figure}[H]%
\centering
\includegraphics[width=0.8\columnwidth]{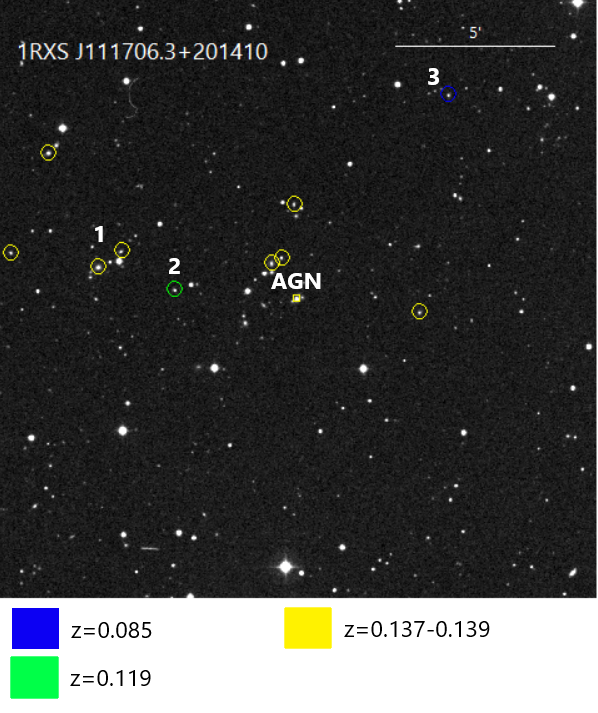} 

(a) Figure 8a
\includegraphics[width=0.8\columnwidth]{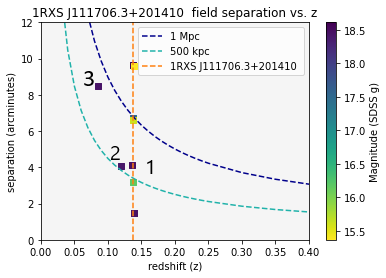} 

(b) Figure 8b
\caption{   }
\label{fig8}
\end{figure}
10 objects were found in a search within 10$^\prime$ of the AGN, all between magnitude 15.4-18.6 (SDSS g). There are 8 objects that are located at the same redshift as the AGN (0.15) in Figure \ref{fig8} (Group 1: 2MASS J11164872+2013412, 2MASX J11173447+2015106, 2MASS J11174694+2015379, 2MASX J11173121+2015417, 2MASS J11170667+2017158, 2MASX J11170982+2015186, 2MASX J11170842+2015296, 2MASX J11174157+2018572). This suggests that these objects form a group with the AGN. The objects in this potential group are at a projected physical distance of less than 1 Mpc from the AGN. 2MASX J11172360+2014247 (Object 2: 11:17:23.6,+20:14:25) located 4.1$^\prime$ E of the AGN with a photo-z redshift 0.12. It has a physical distance of approximately 0.5 Mpc from the AGN and an impact parameter of 633 kpc. 2MASX J11164468+2020557 (Object 3: 11:16:44.7,+20:20:56) is located 8.5$^\prime$ NW of the AGN at photo-z redshift 0.09 and an impact parameter of 945 kpc.

\subsubsection{Objects with unknown redshifts}
\begin{figure}[H]%
\centering
\includegraphics[width= 0.8\columnwidth]{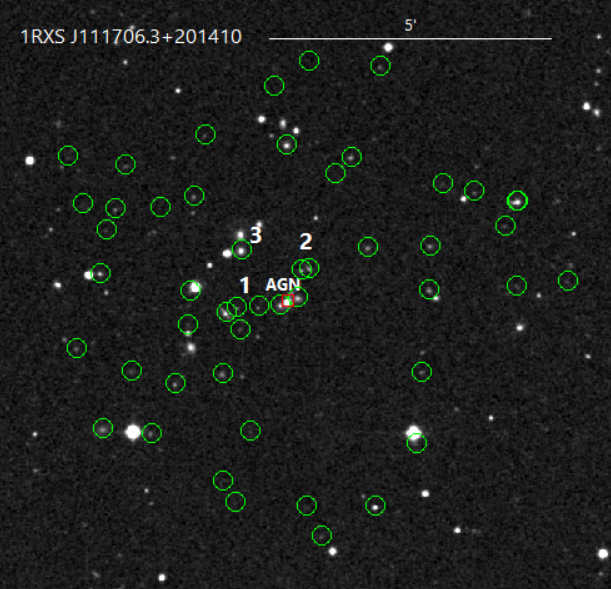}
\caption{There are more than 100 objects in this field, so the search radius was decreased to 5$^\prime$.}
\label{fig9}
\end{figure}
Within a 5$^\prime$ radius of the AGN, there are 55 objects with unknown redshifts. There is a group of 6 objects (Group 1) with an angular separation of  0.15-0.91$^\prime$ E in Figure \ref{fig9} between magnitude 19.9-20.0. There are a pair of objects close to this group (Group 2), located 0.60-0.70$^\prime$ N of the AGN. There is another object (Object 3: 11:17:09.78,+20:15:01.8) of magnitude 18.0 1.2$^\prime$ NE of the AGN. Information on all objects in this field can be found in Table \ref{tbl7}.
\subsection{1RXS J151747.3+652522}
\subsubsection{Objects with known redshifts}
\begin{figure}[H]%
\centering
\includegraphics[width=0.8\columnwidth]{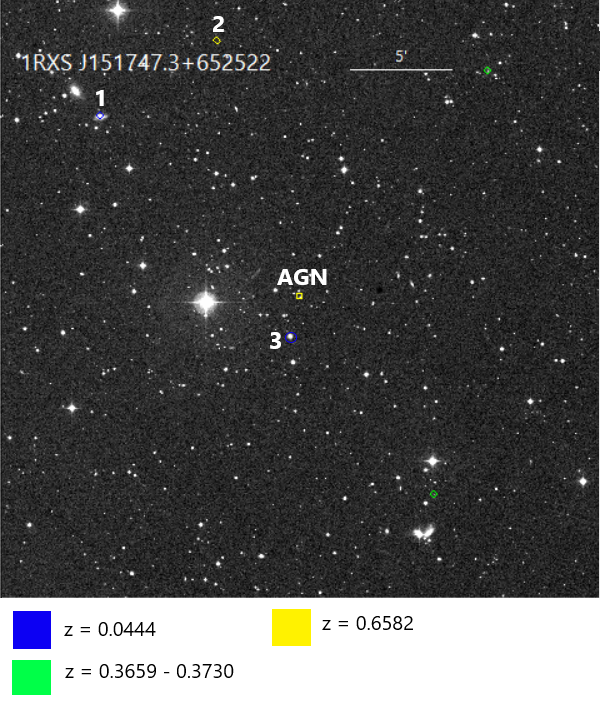} 

(a) Figure 10a
\includegraphics[width=0.8\columnwidth]{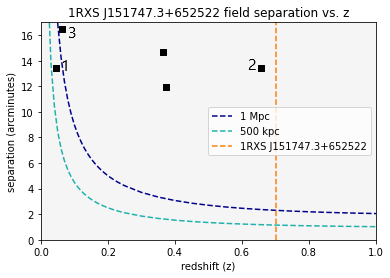} 

(b) Figure 10b
\caption{   (a) Figure 10a has an angular size of 20$^\prime$, as There are no objects with available redshifts less than or equal to 10$^\prime$ from the AGN.(b) There is incomplete optical photometry for these sources, so magnitudes are not plotted in Figure 10b.}
\label{fig10}
\end{figure}
Five objects were found with known redshifts within 15$^\prime$ of the AGN. MCG +11-19-005 (Object 1: 15:19:21.8,+65:34:40) at spectroscopic redshift 0.044 and magnitude 13.6 is located 13.5$^\prime$ NE in Figure \ref{fig10}, corresponding to a physical distance of 0.5 Mpc and impact parameter of 786 kpc. WHL J151824.2+653816 (Object 2: 15:18:24.1,+65:38:16) closest to the redshift of the AGN at photo-z redshift 0.6582, located 13.4$^\prime$ NE of the AGN and greater than 1 Mpc. 2MASX J15175301+6523256 (Object 3: 15:17:53.07,+65:23:26.52) is located 2.0$^\prime$ SW at magnitude 16.5, physical distance of 1 Mpc, and photo-z redshift 0.062.
\subsubsection{Objects with unknown redshifts}
\begin{figure}[H]%
\centering
\includegraphics[width= 0.8\columnwidth]{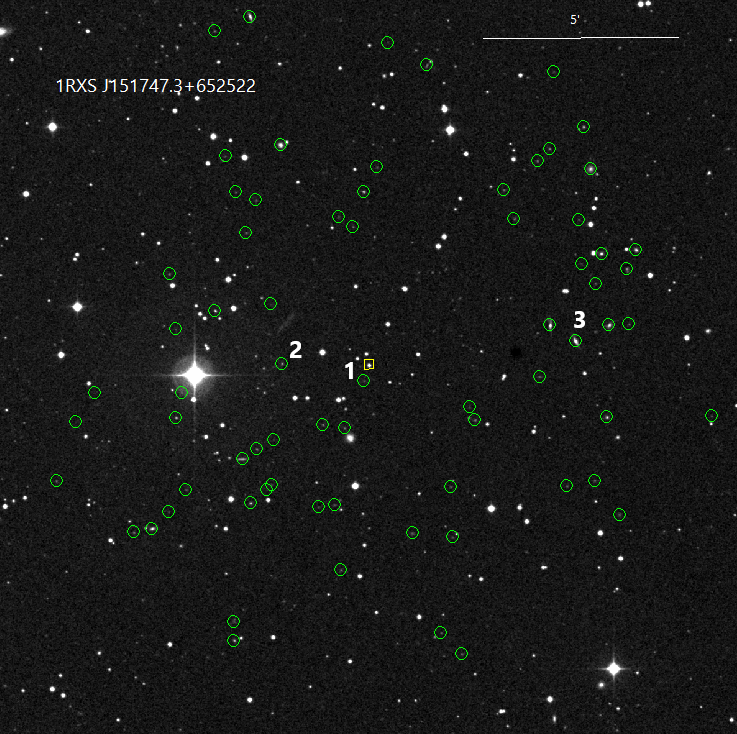}
\caption{   }
\label{fig11}
\end{figure}
There are 93 objects between 16.5 and 20.9 magnitude with unknown redshifts found within 10$^\prime$ of the AGN. The closest in separation is WISEA J151748.75+652456.8 (Object 1: 15:17:48.93,+65:24:58.0) located 0.45$^\prime$ SE of the AGN in Figure \ref{fig11} at magnitude 20.4. WISEA J151810.21+652529.7 (Object 2: 15:18:10.2,+65:25:29.68) is located 2.4$^\prime$ NE of the AGN at magnitude 20.3. There are two objects (WISEA J151700.10+652618.1 and 2MASX J15165360+6525518, Group 3) located 5.0$^\prime$ W of the AGN at magnitudes 17.6 and 17.3 respectively. Information on all objects in this field can be found in Table \ref{tbl8}.
\subsection{1RXS J153501.1+532042}
\subsubsection{Objects with known redshifts}
\begin{figure}[H]%
\centering
\includegraphics[width=0.8\columnwidth]{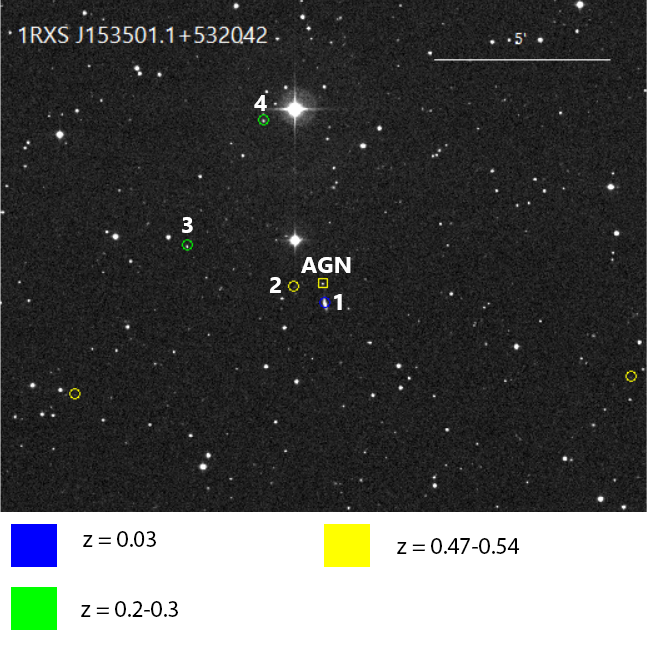} 

(a) Figure 12a
\includegraphics[width=0.8\columnwidth]{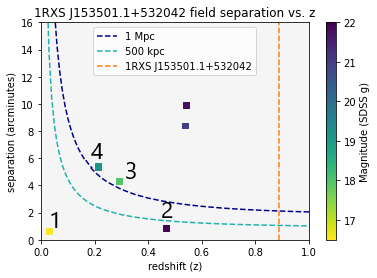} 

(b) Figure 12b
\caption{   }
\label{fig12}
\end{figure}
There are six objects with known redshifts found within 10$^\prime$ of the AGN. WISEA J153500.26+532000.7 (Object 1: 15:35:00.2,+53:20:00) is the closest object at spectroscopic redshift 0.029 and magnitude 16.5, located 0.63$^\prime$ S from the AGN in Figure \ref{fig12} corresponding to a physical distance of less than 0.5 Mpc. It has an estimated impact parameter of  24.9 kpc. The next closest object is WISEA J153506.70+532029.2 (Object 2: 15:35:06.7,+53:20:29) at spectroscopic redshift 0.47 and magnitude 22, with separation 0.89$^\prime$ SE corresponding to less than 0.5 Mpc. It has an impact parameter of 537 kpc. SBS 1534+53 (Object 3: 15:35:29.1,+53:21:39) at spectroscopic redshift 0.29 and magnitude 18.0 is located 4.35$^\prime$ NE at approximately 1 Mpc, with an impact parameter of 1630 kpc. WISEA J153514.10+532535.8 (Object 4: 15:35:14.1,+53:25:36) at spectroscopic redshift 0.21 and magnitude 19.2 is located 5.36$^\prime$ NE of the AGN. It is at a physical distance of 1 Mpc with an impact parameter of 1460 kpc.
\subsubsection{Objects with unknown redshifts}
\begin{figure}[H]%
\centering
\includegraphics[ width=0.8\columnwidth]{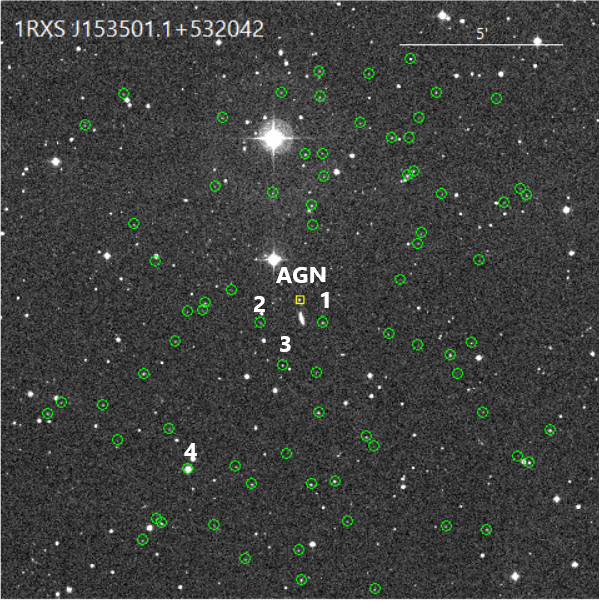}
\caption{   }
\label{fig13}
\end{figure}
There are 76 objects with unknown redshifts found within 10$^\prime$ of the AGN. Closest in separation is WISEA J153455.37+531952.6 (Object 1: 15:34:55.39,+53:19:52.61) at magnitude 19.8 located 1.1$^\prime$ S of the AGN in Figure \ref{fig13}. WISEA J153509.26+531948.7 (Object 2: 15:35:9.27,+53:19:48.9) at magnitude 20.3 is located 1.5$^\prime$ SE of the AGN. WISEA J153504.05+531825.2 (Object 3: 15:35:4.06,+53:18:25.16) at magnitude 20.2 is located 2.6$^\prime$ S of the AGN. WISEA J153519.25+532634.8 (Object 4: 15:35:19.24,+53:26:35.56) is the brightest object in the field at magnitude 14.5, located 6.57$^\prime$ SE of the AGN. Information on all objects in this field can be found in Table \ref{tbl9}.
\subsection{3C 279}
\subsubsection{Objects with known redshifts}
\begin{figure}[H]%
\centering
\includegraphics[width=0.8\columnwidth]{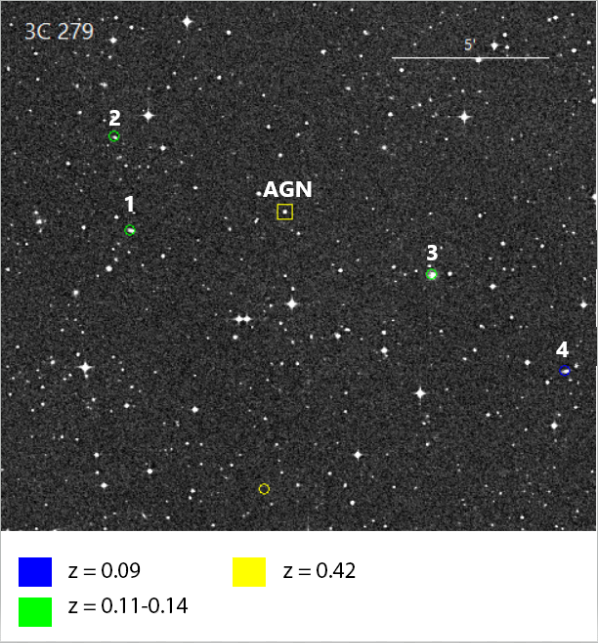} 

(a) Figure 14a
\includegraphics[width=0.8\columnwidth]{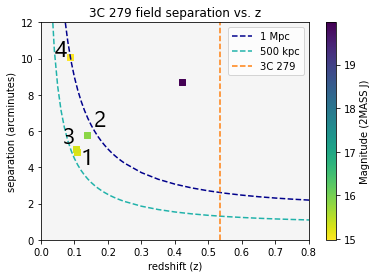} 

(b) Figure 14b
\caption{   (a) The size of the image was reduced to 10x10$^\prime$ to obtain better resolution on the nearest objects to the AGN. Six objects were cut from the image, all located at a separation of 8$^\prime$ or greater.}
\label{fig14}
\end{figure}
There are 5 objects found within 10$^\prime$ of the AGN with known redshifts in Figure \ref{fig14}. LCRS B125355.2-053144 (Object 1: 12:56:30.5,-05:47:58) at spectroscopic redshift 0.11 and magnitude 15.2 (2MASS J) is located 4.85$^\prime$ E of the AGN, corresponding to 0.5 Mpc. LCRS B125357.3-052850 (Object 2: 12:56:32.6,-05:45:04) at spectroscopic redshift 0.14 and magnitude 15.8 is 5.02$^\prime$ NE of the AGN. This corresponds to a physical distance of less than 1 Mpc and an impact parameter of 10420kpc. LCRS B125317.3-053305
(Object 3: 12:55:52.6,-05:49:20) at spectroscopic redshift 0.11 has angular separation 5.0$^\prime$ SW and magnitude 15.3, and projected physical distance of around 0.5 Mpc. This object has an impact parameter of 697 kpc. LCRS B125300.5-053604 (Object 4: 12:55:35.8,-05:52:19) at spectroscopic redshift 0.09 is separated 10$^\prime$ SW of the AGN at magnitude 15.0. This correspons to a physical ditance of less than 0.5 Mpc.
\subsubsection{Objects with unknown redshifts}
\begin{figure}[H]%
\centering
\includegraphics[width= 0.8\columnwidth]{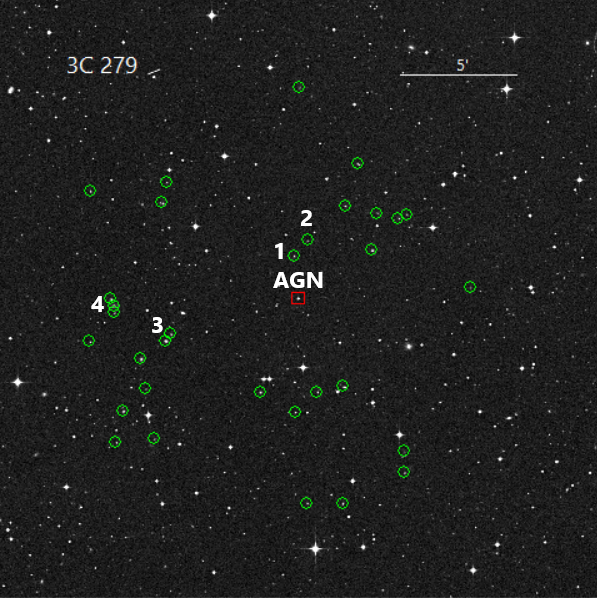}
\caption{   }
\label{fig15}
\end{figure}
There are 33 objects found within 10$^\prime$ of the AGN with unknown redshifts. The closest in separation is WISEA J125611.81-054535.6 (Object 1: 12:56:11.86,-05:45:35.75) at magnitude 19.8, located 1.8$^\prime$ N of the AGN in Figure \ref{fig15}. Next closest is WISEA J125609.77-054458.7 (Object 2: 12:56:9.59,-05:44:56.98) at magnitude 19.55 located 1.8$^\prime$ N. A possible group of two closely separated objects is found approximately 7.0$^\prime$ SE of the AGN (Object 3). Finally, there is another possible group of three objects (Object 4) located 7.7$^\prime$ E of the AGN. Information on all objects in this field can be found in Table \ref{tbl10}. 

\subsection{3C 454.3}
\subsubsection{Objects with known redshifts}

\begin{figure}[H]%
\centering
\includegraphics[width=0.8\columnwidth]{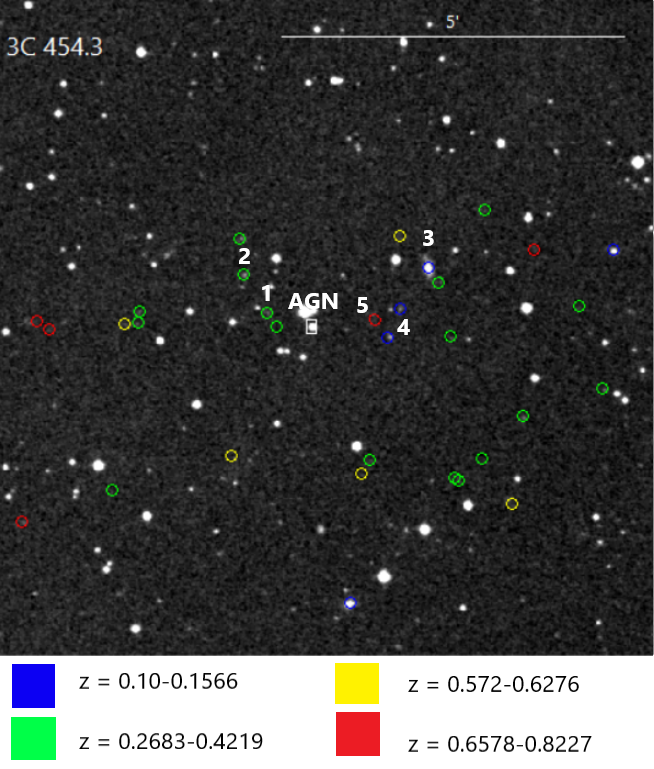} 

(a) Figure 16a
\includegraphics[width=0.8\columnwidth]{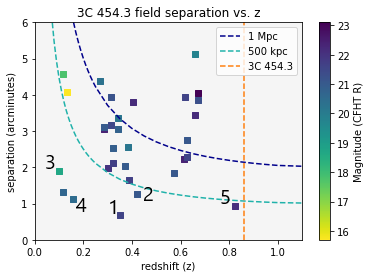} 

(b) Figure 16b
\caption{   Angular size of the figure was reduced from 20$^\prime$ to 10$^\prime$ to reduce the number of objects. }
\label{fig16}
\end{figure}
There are 33 objects found within 6$^\prime$ of the AGN with known redshifts in Figure \ref{fig16}. SSTSL2 J225400.41+160906.4 (Object 1: 22:54:0.5,+16:09:6.01) at spectroscopic redshift 0.35 and magnitude 21.6 (CFHT R) is located 0.693$^\prime$ E of the AGN, corresponding to a physical distance less than 0.5 Mpc. SSTSL2 J225401.86+160939.6 (Object 2: 22:54:1.9,+16:09:38.99) at spectroscopic redshift 0.42 and magnitude 21.1 is 1.25$^\prime$ NE of the AGN, with a physical distance close to 0.5 Mpc. 2MASX J22535061+1609434 (Object 3: 22:53:50.6,+16,09,43.6) is the lowest redshift object in the field, at specroscopic redshift 0.10 and magnitude 18.7. It is located 1.9$^\prime$ NW of the AGN at a physical distance less than 0.5 Mpc, with an impact parameter of 244 kpc. SSTSL2 J225352.98+160843.9 (Object 4: 22:53:53.1,+16:08:43.01) at spectroscopic redshift 0.157 and magnitude 20.5 is located 1.1$^\prime$ W of the AGN with physical distance less than 0.5 Mpc. Lastly, SSTSL2 J225353.90+160858.4 (Object 5: 22:53:53.9,+16:08:57.98) at spectroscopic redshift 0.823 and magnitude 22.1 is separated 0.93$^\prime$ W, with a physical distance of about 0.5 Mpc.

\subsubsection{Objects with unknown redshifts}
\begin{figure}[H]%
\centering
\includegraphics[ width=0.8\columnwidth]{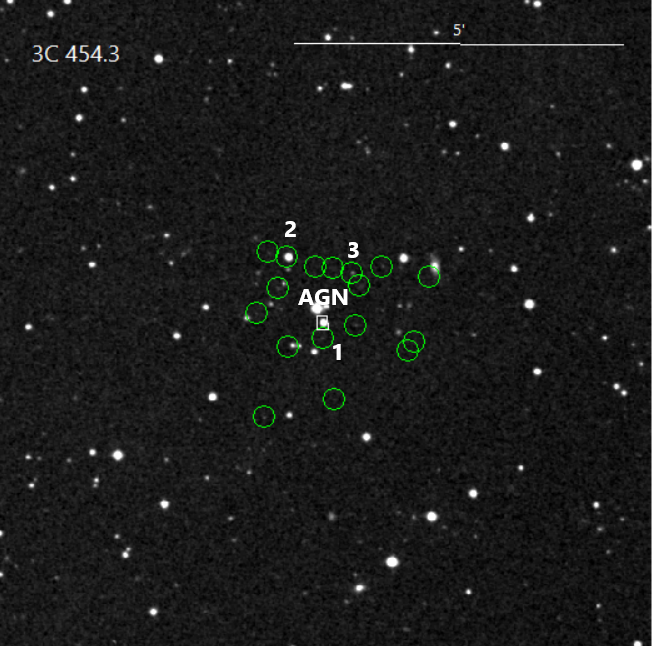}
\caption{   }
\label{fig17}
\end{figure}
There are a total of 2,190 objects found within 10$^\prime$ including unclassified gamma-ray, x-ray, infrared and visual sources. When the list was narrowed down to only include galaxies and visual sources, the result was 18 objects of interest within 10$^\prime$ of the AGN. The first object is an unclassified visual source 3C 454.3:[KSS94] 28 (Object 1: 22:53:57.8,+16:08:39.01) at magnitude 23.1 is located 0.243$^\prime$ S of the AGN in Figure \ref{fig17}. WISEA J225359.94+160953.9 (Object 2: 22:54:0.0,+16:09:52.99) at magnitude 16.2 is located 1.13$^\prime$ NE of the AGN. WISEA J225355.89+160938.9 (Object 3: 22:53:55.9,+16:09:38.02) at magnitude 21.0 is located 0.86$^\prime$ NW of the AGN. Information on all objects in this field can be found in Table \ref{tbl11}.

\subsection{H2356-309}
\subsubsection{Objects with known redshifts}
\begin{figure}[H]%
\centering
\includegraphics[width=0.8\columnwidth]{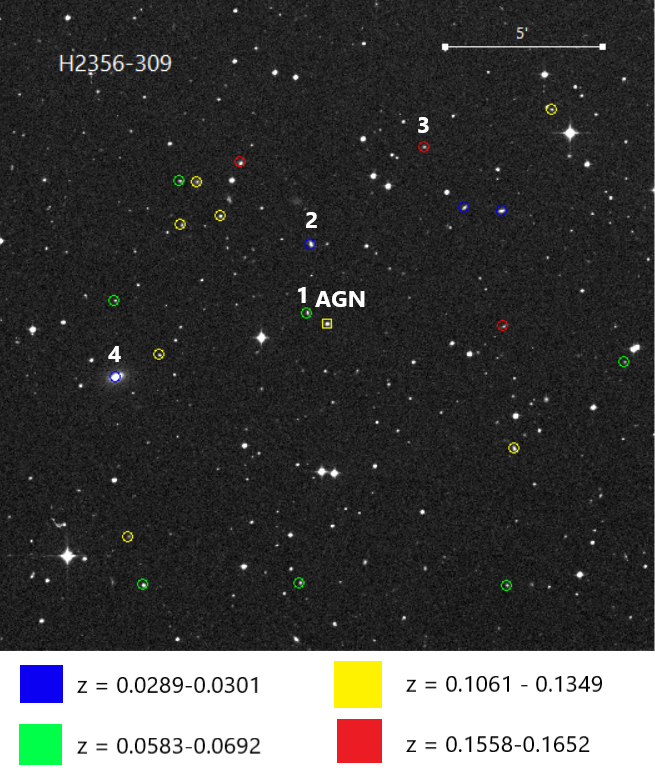} 

(a) Figure 18a
\includegraphics[width=0.8\columnwidth]{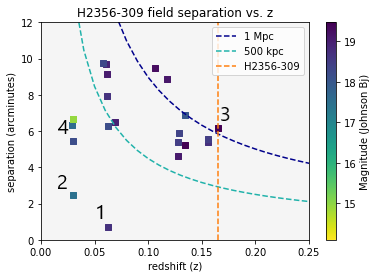} 

(b) Figure 18b
\caption{   }
\label{fig18}
\end{figure}
There are a total of 24 objects found within 10$^\prime$ of the AGN. WISEA J235910.71-303720.6 (Object 1: 23:59:10.73,-30:37:21.61) at spectroscopic redshift 0.062 and magnitude 18.7 is located 0.68$^\prime$ NE of the AGN in Figure \ref{fig18}, corresponding to a physical distance of less than 0.5 Mpc. WISEA J235910.27-303514.7 (Object 2:23:59:10.32,-30:35:15.61) at spectroscopic redshift 0.029 and magnitude 17.5 is 2.47$^\prime$ N, also at a physical distance of less than 0.5 Mpc. WISEA J235854.35-303214.9 (Object 3: 23:58:54.35,-30:32:15.9) at spectroscopic redshift 0.165 and magnitude 19.5 is located 6.15$^\prime$ NW and at a physical distance of approximately 1 Mpc. 6dF J2359377-303921 (Object 4: 23:59:37.8,-30:39:20.92) at spectroscopic redshift 0.030 and magnitude 14.1 (SDSS r) is located 6.64$^\prime$ SE of the AGN. It has a physical distance of less than 0.5 Mpc.

\subsubsection{Objects with unknown redshifts}
\begin{figure}[H]%
\centering
\includegraphics[width= 0.8\columnwidth]{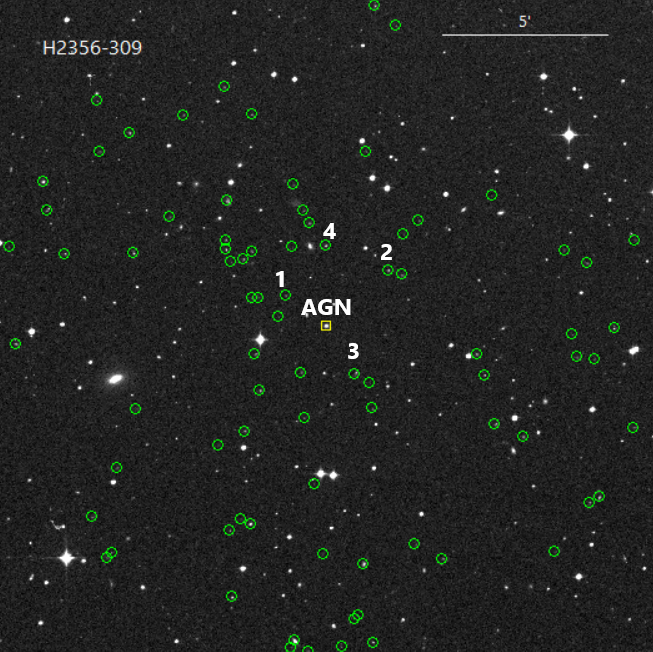}
\caption{   }
\label{fig19}
\end{figure}
There are a total of 99 objects with unavailable redshifts found within 10$^\prime$ of the AGN. WISEA J235913.66-303645.8 (Object 1: 23:59:13.67,-30:36:45.97) at magnitude 20.5 is located 1.54$^\prime$ NE in Figure \ref{fig19}. WISEA J235859.11-303557.9 (Object 2: 23:58:59.14,-30:35:58.09) at magnitude 19.9 is 2.546$^\prime$ NW of the AGN. WISEA J235903.56-303906.3 (Object 3: 23:59:3.71,-30:39:8.28) at magnitude 19.4 is located 1.72$^\prime$ SW. WISEA J235908.04-303513.8 (Object 4: 23:59:8.04,-30:35:13.99) at magnitude 18.3 (SDSS r) is 2.44$^\prime$ N. Information on all objects in this field can be found in Table \ref{tbl12}.
\subsection{S5 0836+71}
\subsubsection{Objects with known redshifts}
\begin{figure}[H]%
\centering
\includegraphics[width=0.8\columnwidth]{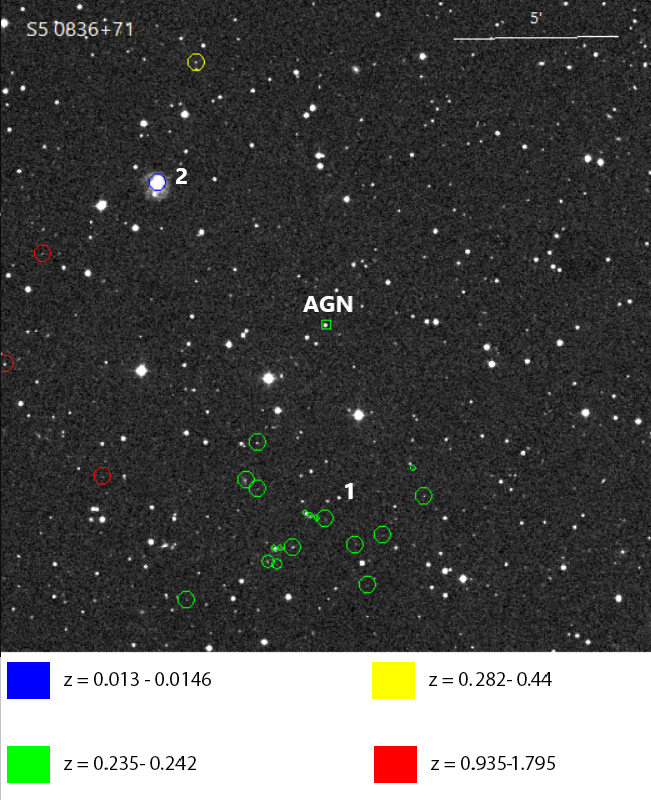} 

(a) Figure 20a
\includegraphics[width=0.8\columnwidth]{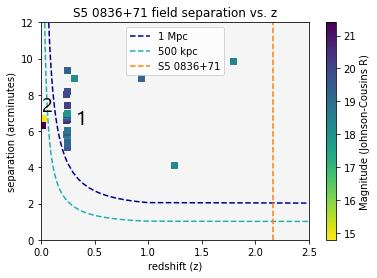} 

(b) Figure 20b
\caption{   }
\label{fig20}
\end{figure}
There are 31 objects with known redshifts found within 10$^\prime$ of the AGN. There is a large grouping of at least 22 galaxies at redshift 0.24 located approximately 6$^\prime$ S of the AGN (Group 1) in Figure \ref{fig20}, greater than 1 Mpc in projected physical distance. UGC 04522 (Object 2: 08:42:26.6,+70:58:05) at spectroscopic redshift 0.015, physical distance less than 0.5 Mpc, and magnitude 14.80 is located 6.7$^\prime$ NE and has an impact parameter of 133 kpc. 
\subsubsection{Objects with unknown redshifts}
\begin{figure}[H]%
\centering
\includegraphics[ width=0.8\columnwidth]{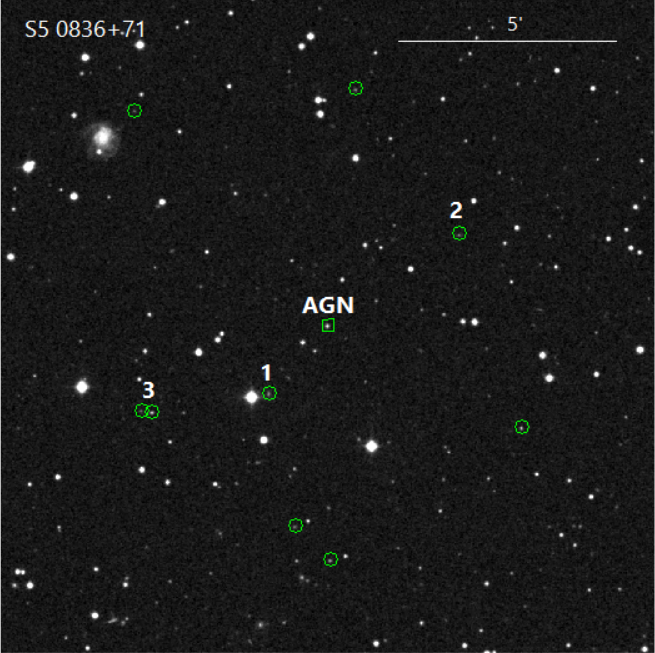}
\caption{   Angular size of the image was reduced from 20$^\prime$ to 15$^\prime$ due to there being more than 100 objects returned in a 10$^\prime$ search radius.}
\label{fig21}
\end{figure}
There are a total of 10 objects with unavailable redshifts found within 10$^\prime$. WISEA J084140.99+705210.4 (Object 1: 08:41:40.84,+70:52:9.98) at magnitude 18.9 is located 2.04$^\prime$ SE of the AGN in Figure \ref{fig21}. WISEA J084047.07+705544.2 (Object 2: 08:40:46.95,+70:55:45.01) at magnitude 19.24 is located 3.68$^\prime$ NW. WISEA J084216.83+705148.6 and WISEA J084030.58+705117.9 (Group 3) at magnitudes 19.3 and 18.3 respectively are located 5.0$^\prime$ SE. Information on all objects in this field can be found in Table \ref{tbl13}.

\subsection{Ton 116}
\subsubsection{Objects with known redshifts}
\begin{figure}[H]%
\centering
\includegraphics[width=0.8\columnwidth]{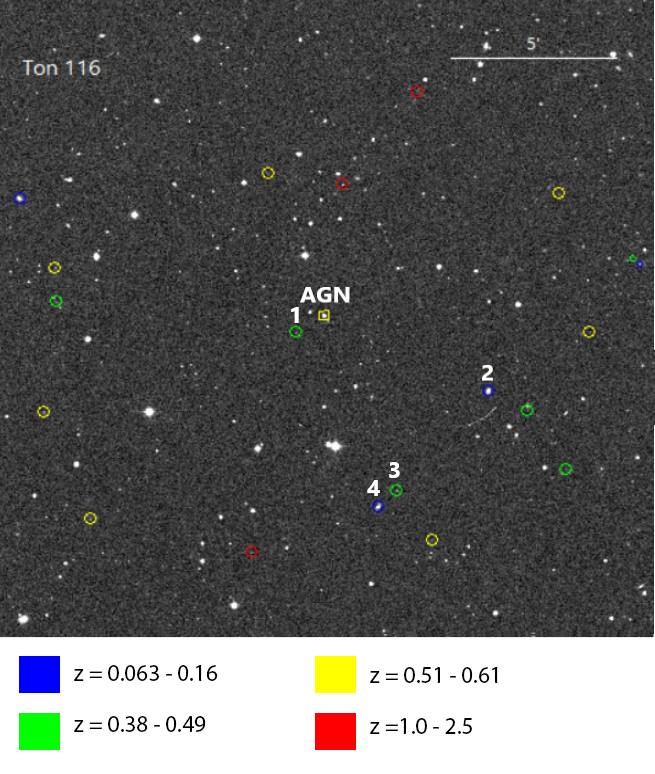} 

(a) Figure 22a
\includegraphics[width=0.8\columnwidth]{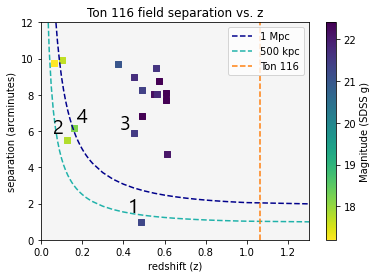} 

(b) Figure 22b
\caption{   }
\label{fig22}
\end{figure}
There are 22 objects with known redshifts found within 10$^\prime$ of the AGN. WISEA J124316.96+362713.6 (Object 1: 12:43:16.97,+36:27:13.79) at spectroscopic redshift 0.485, physical distance less than 0.5 Mpc, and magnitude 21.3 (SDSS g) is located 0.99$^\prime$ SE of the AGN in Figure \ref{fig22}. This corresponds to an impact parameter of 618 kpc. WISEA J124248.22+362517.6 (Object 2: 12:42:48.24,+36:25:17.72) at spectroscopic redshift 0.128 and magnitude 17.8 (SDSS g) is located 5.5$^\prime$ SW, corresponding to a physical distance less than 1 Mpc and an impact parameter of 915 kpc. WISEA J124302.14+362213.4 (0bject 3: 12:43:2.18,+36:22:13.69) at spectroscopic redshift 0.452, physical distance greater than 1 Mpc, and magnitude 21.5 (SDSS g) is located 5.9$^\prime$ S. This object has an impact parameter of 3440 kpc. WISEA J124305.07+362144.1 (Object 4: 12:43:5.08,+36:21:44.17) at spectroscopic redshift 0.158 and magnitude 18.2 (SDSS g) is located 6.19$^\prime$ S of the AGN corresponding to a physical distance of approximately 1 Mpc. This object has an impact parameter of 1270 kpc.
\subsubsection{Objects with unknown redshifts}
\begin{figure}[H]%
\centering
\includegraphics[ width=0.8\columnwidth]{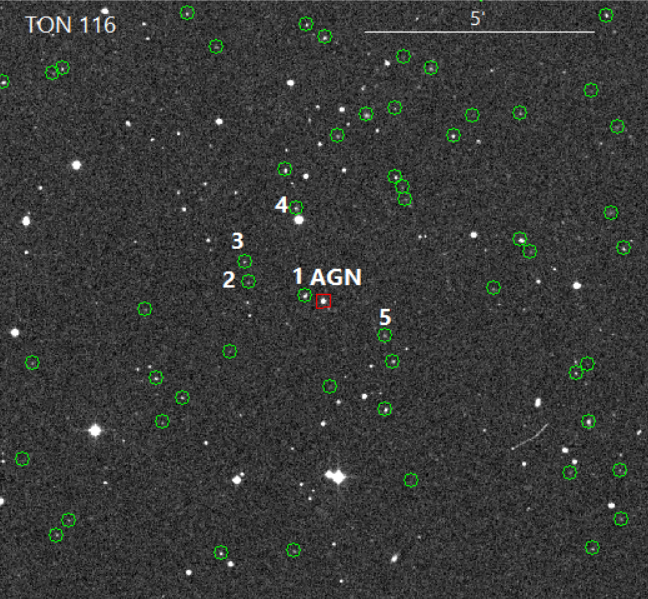}
\caption{   }
\label{fig23}
\end{figure}
There are 100 objects with unavailable redshifts within 10$^\prime$ of the AGN. Redshifts for more of these objects would be vital, as there are several bright objects in the field that could be possible absorption hosts besides the 5 objects indicated in Figure \ref{fig23}. WISEA J124314.74+362752.6 (Object 1: 12:43:14.7,+36:27:53.35) at magnitude 20.6 is located 0.424$^\prime$ E of the AGN. WISEA J124321.34+362810.8 (Object 2: 12:43:21.33,+36:28:11.03) at magnitude 20.8 is located 1.79$^\prime$ E of the AGN. WISEA J124321.71+362839.8 (Object 3: 12:43:21.72,+36:28:39.9) at magnitude 20.8 is located 2.03$^\prime$ NE of the AGN. WISEA J124315.69+362952.9 (Object 4: 12:43:15.72,+36:29:53.27) at magnitude 19.6 is located 2.23$^\prime$ N of the AGN. WISEA J124305.68+362654.3 (Object 5: 12:43:5.7,+36:26:54.2) at magnitude 20.8 is located 1.64$^\prime$ SW. Information on all objects in this field can be found in Table \ref{tbl14}.

\subsection{1RXS J122121.7+301041}
\subsubsection{Objects with known redshifts}
\begin{figure}[H]%
\centering
\includegraphics[width=0.8\columnwidth]{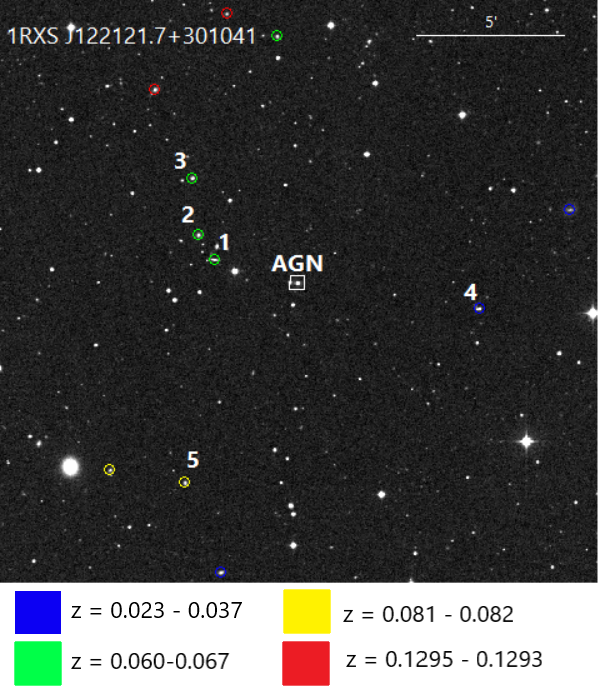} 

(a) Figure 24a
\includegraphics[width=0.8\columnwidth]{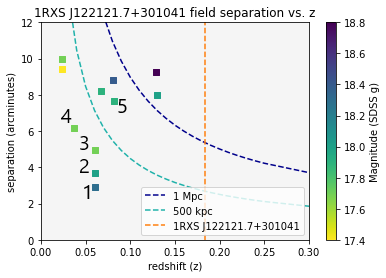} 

(b) Figure 24b
\caption{   }
\label{fig24}
\end{figure}
Within a radius of 10$^\prime$ around the AGN, There are 11 sources identified with available redshifts in Figure \ref{fig24}. The closest to the AGN is WISEA J122134.81+301123.1 (Object 1: 12:21:34.84,+30:11:23.39) at spectroscopic redshift 0.060, physical distance less than 0.5 Mpc, and with magnitude 18.3 (SDSS g) located 3.0$^\prime$ NE. This corresponds to a physical distance of less than 0.5 Mpc and an impact parameter of 230 kpc. 3.6$^\prime$ NE of the AGN is WISEA J122137.29+301211.7 (Object 2: 12:21:37.29,+30:12:11.77) also at spectroscopic redshift 0.06 and with magnitude 18. This corresponds to a physical distance less than 0.5 Mpc and an impact parameter of 291 kpc. WISEA J122138.25+301405.3 (Object 3: 12:21:38.25,+30:14:5.5) at spectroscopic redshift 0.06 and magnitude 17.7 is 4.9$^\prime$ NE of the AGN. This object is at a physical distance of less than 0.5 Mpc and has an impact parameter of 394 kpc. These three objects are all at the same redshift, so it is possible that they comprise a galaxy group.
SDSS J122053.73+300947.2 (Object 4: 12:20:53.73,+30:09:47.2) is at spectroscopic redshift 0.04 and magnitude 17.7, located 6.1$^\prime$ W. This corresponds to a physical distance of less than 0.5 Mpc. There are also two objects at a redshift of 0.08, the closest in angular separation being WISEA J122139.10+300356.7 (Object 5: 12:21:39.11,+30:03:56.88). The object is at magnitude 17.9 and 7.6$^\prime$ SE iwth an impact parameter of 818 kpc and physical distance less than 1 Mpc. 
\subsubsection{Objects with unknown redshifts}
\begin{figure}[H]%
\centering
\includegraphics[width= 0.8\columnwidth]{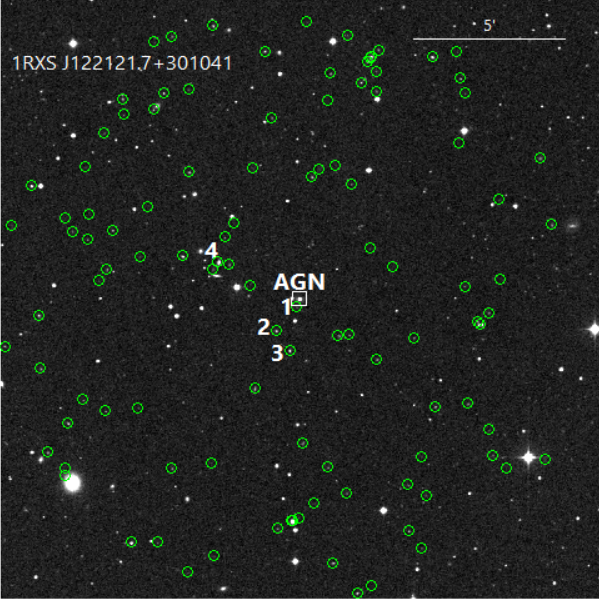}
\caption{   }
\label{fig25}
\end{figure}
There are a total of 108 objects below 21 mag found within 10$^\prime$ of the AGN. WISEA J122122.35+301021.1 (Object 1: 12:21:22.34,+30:10:20.78) at magnitude 20.6 (SDSS g) has the least angular separation of 0.28$^\prime$ SE in Figure \ref{fig25}.
WISEA J122125.49+300932.7 (Object 2: 12:21:25.49,+30:09:32.87) is at magnitude 19.6 and separation 1.3$^\prime$ SE. WISEA J122123.38+300853.6 (Object 3: 12:21:23.39,+30:08:53.7) is at magnitude 19.2 and separation 1.8$^\prime$ S. WISEA J122134.44+301149.1 (Object 4: 12:21:34.48,+30:11:49.13) at magnitude 18.5 is located 2.9$^\prime$ NE from the AGN. Information on all objects in this field can be found in Table \ref{tbl15}.
\subsection{1RXS J142239.1+580159}
\subsubsection{Objects with known redshifts}
\begin{figure}[H]%
\centering
\includegraphics[width=0.8\columnwidth]{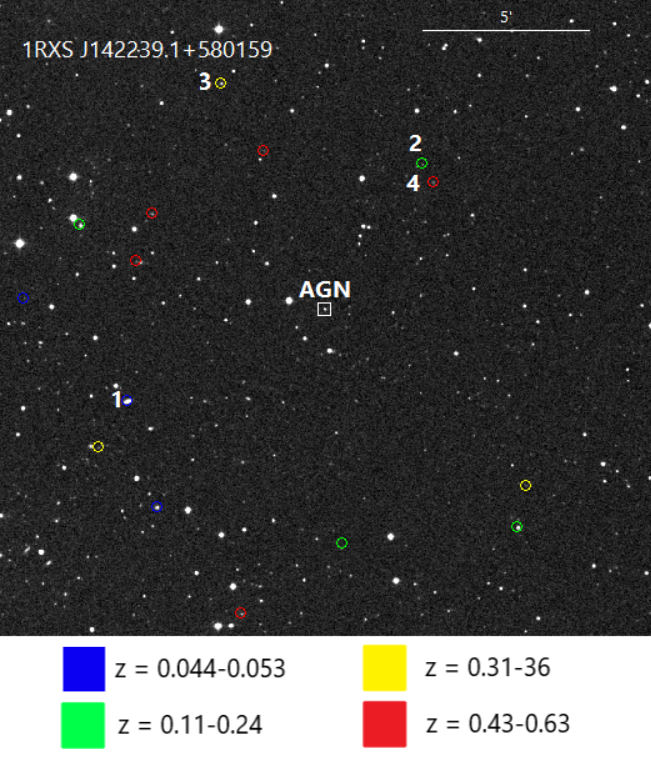} 

(a) Figure 26a
\includegraphics[width=0.8\columnwidth]{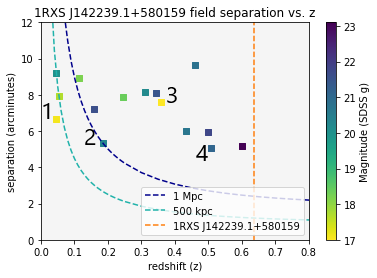} 

(b) Figure 26b
\caption{   }
\label{fig26}
\end{figure}
Within a radius of 10$^\prime$ around the AGN, we found 17 objects. WISEA J142325.43+575923.7 (Object 1: 14:23:25.43,+57:59:23.5) at spectroscopic redshift 0.044 and magnitude 17.1 (SDSS g) has an angular separation of 6.7$^\prime$ SE of the AGN in Figure \ref{fig26}, and a corresponding physical distance of around 0.5 Mpc. This object has an impact parameter of 388 kpc. The next candidate, SDSS J142214.68+580610.9 (Object 2: 14:22:14.69,+58:06:10.94) at spectroscopic redshift 0.18 and magnitude 20.3 is 5.3$^\prime$ NW. This corresponds to a physical distance of 1 Mpc. WISEA J142300.36+580856.9 (Object 3: 14:23:0.43,+58:08:56.94) at spectroscopic redshift 0.36 and magnitude 17.0 is 7.6$^\prime$ N. This object has a physical distance greater than 1 Mpc. Finally, WISEA J142212.41+580537.1 (Object 4: 14:22:12.47,+58:05:37.64) is at spectroscopic redshift 0.51 and magnitude 21.1 and 5.1 $^\prime$ NW. This corresponds to a physical distance greater than 1 Mpc and an impact parameter of 1940 kpc.
\subsubsection{Objects with unknown redshifts}
\begin{figure}[H]%
\centering
\includegraphics[ width=0.8\columnwidth]{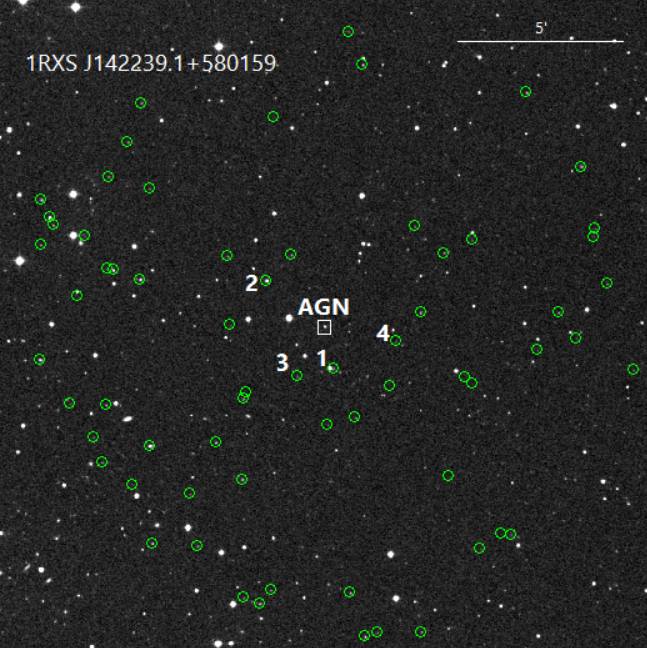}
\caption{   }
\label{fig27}
\end{figure}
We found 66 objects below 21 mag (SDSS g). The first object is SDSS J142237.38+580036.1 (Object 1: 14:22:37.38,+58:00:36.14) at magnitude 20 and separation 1.3$^\prime$ S in Figure \ref{fig27}. WISEA J142251.82+580324.2 (Object 2: 14:22:51.85,+58:03:24.19) is at magnitude 19.1 and is 2.2$^\prime$ NE. WISEA J142245.79+580026.4 (Object 3: 14:22:45.78,+58:00:26.28) is at magnitude 20.8 and located 1.7$^\prime$ SE. WISEA J142222.64+580122.1 (Object 4: 14:22:22.7,+58:01:21.18) is at magnitude 20.5 and 2.2$^\prime$ SW. Information on all objects in this field can be found in Table \ref{tbl16}.
\subsection{1RXS J150759.8+041511}
\subsubsection{Objects with known redshifts}
\begin{figure}[H]%
\centering
\includegraphics[width=0.8\columnwidth]{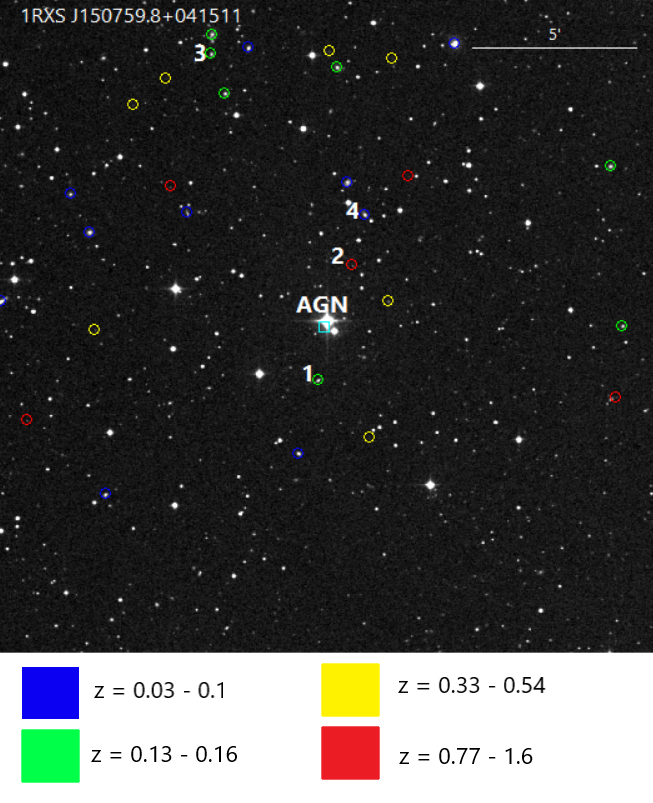} 

(a) Figure 28a
\includegraphics[width=0.8\columnwidth]{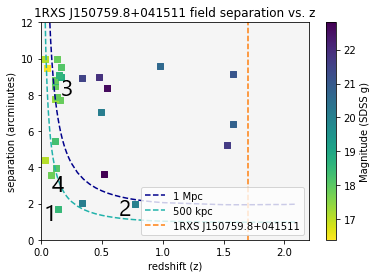} 

(b) Figure 28b
\caption{   }
\label{fig28}
\end{figure}
Within a radius of 10$^\prime$ around the AGN, we found 30 objects. WISEA J150800.69+041332.0 (Object 1: 15:08:0.7,+04:13:32.02) at spectroscopic redshift 0.14 and magnitude 18.5 (SDSS g) is located 1.7$^\prime$ S of the AGN in Figure \ref{fig28}. This corresponds to a physical distance less than 0.5 Mpc and an impact parameter of 307 kpc. WISEA J150756.46+041701.4 (Object 2: 15:07:56.45,+04:17:1.64) is at spectroscopic redshift 0.78 and magnitude 20.0 is 2.0$^\prime$ N. This object has a physical distance less than 1 Mpc and an impact parameter of 1010 kpc. There appears to be a grouping of 5 galaxies at redshifts of 0.13-0.16 (Group 3) about 9$^\prime$ NE. WISEA J150754.93+041834.7 (Object 4: 15:07:54.94,+04:18:34.7) has a spectroscopic redshift 0.081 and magnitude 17.8 is 3.6$^\prime$ N of the AGN. This corresponds to a physical distance of less than 0.5 Mpc and an impact parameter of 383 kpc.
\subsubsection{Objects with unknown redshifts}
\begin{figure}[H]%
\centering
\includegraphics[width= 0.8\columnwidth]{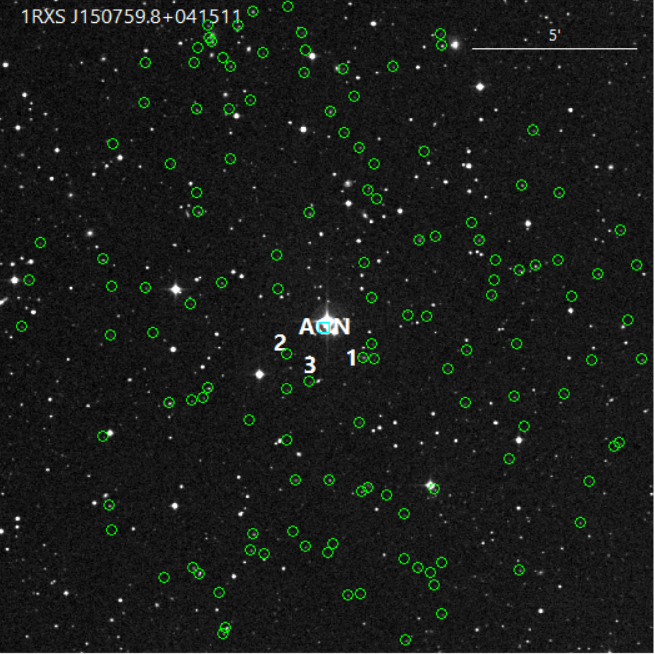}
\caption{   }
\label{fig29}
\end{figure}
We found 133 objects below 21 mag (SDSS g). First is SDSS J150755.09+041414.1 (Object 1: 15:07:55.09,+04:14:14.14) at magnitude 19 and separation 1.5$^\prime$ S in Figure \ref{fig29}. WISEA J150804.48+041420.7 (Object 2: 15:08:4.51,+04:14:21.05) is at magnitude 20.0 and 1.4$^\prime$ SE. WISEA J150801.68+041328.6
(Object 3: 15:08:1.69,+04:13:28.7) at magnitude 20.8 has a separation of 1.8$^\prime$ SE of the AGN. Information on all objects in this field can be found in Table \ref{tbl17}.

\begin{deluxetable*}{lllllll}
\tablenum{2}
\label{tbl2}
\tablecaption{Target Fields Needing Deeper Imaging and Spectroscopy}
\tablewidth{0pt}
\tablehead{ \colhead{AGN} & \colhead{RA (h:m:s)}  & \colhead{DEC (d:m:s)}  & \colhead{Redshift}  & \colhead{No. known} & \colhead{No. unknown} & \colhead{$<$ 0.5 Mpc}
}
\startdata
1RXS   J003334.6-192130 & 00:33:34.200  & -19:21:33.30  & 0.610 & 0  & 108 & 0 \\
1RXS J022716.6+020154   & 02:27:16.580  & +02:02:00.50  & 0.457 & 12 & 25  & 2 \\
S5 0836+71              & 08:41:24.3652 & +70:53:42.173 & 2.172 & 31 & 10  & 2 \\
1ES   1028+511          & 10:31:18.518  & +50:53:35.82  & 0.360 & 5  & 35  & 3 \\
1RXS   J110337.7-232931 & 11:03:37.610  & -23:29:31.20  & 0.186 & 2  & 4   & 2 \\
1RXS   J111706.3+201410 & 11:17:06.260  & +20:14:07.40  & 0.139 & 10 & 55  & 4 \\
1RXS   J122121.7+301041 & 12:21:21.941  & +30:10:37.11  & 0.184 & 11 & 108 & 6 \\
Ton 116                 & 12:43:12.7362 & +36:27:43.999 & 1.065 & 22 & 100 & 1 \\
3C 279                  & 12:56:11.1665 & -05:47:21.523 & 0.536 & 5  & 33  & 2 \\
1RXS J142239.1+580159   & 14:22:38.895  & +58:01:55.50  & 0.635 & 17 & 66  & 3 \\
1RXS J150759.8+041511   & 15:07:59.7324 & +04:15:11.984 & 1.703 & 30 & 133 & 8 \\
1RXS J151747.3+652522   & 15:17:47.600  & +65:25:23.90  & 0.702 & 4  & 93  & 0 \\
1RXS J153501.1+532042   & 15:35:00.801  & +53:20:37.33  & 0.890 & 6  & 76  & 2 \\
3C 454.3                & 22:53:57.7479 & +16:08:53.560 & 0.859 & 33 & 18  & 9
\enddata
\tablecomments{No. known refers to the number of objects in the field with available redshifts and No. unknown refers to the number without available redshifts. $<$ 0.5 Mpc refers to the number of objects with known redshifts at a physical distance less than 0.5 Mpc. This column, in addition to merit in Table \ref{table1}, is meant to guide observers while prioritizing which background targets to focus on.}
\end{deluxetable*}

\section{Discussion}

The lists of identified objects become incomplete when $z \gtrsim 0.4$ due to the magnitude limit several surveys used such as the SDSS (22.0), when considering target galaxies brighter than 0.25L*; this applies to 10/19 fields. 
Imaging and spectroscopy of the 10 fields around $z > 0.4$ AGN should be followed up with deeper imaging plus spectroscopy, which is possible with 8m class telescopes. 
The number of objects that need spectroscopy per field makes this an ideal project for multi-fiber spectrographs that have become common. We identify fields that need followup imaging and spectroscopy in Table \ref{tbl2}.

The results of this study can be applied to guide the process of matching galaxies and groups to their relevant absorption lines. To narrow possible matches, we can take velocity measurements of likely absorption hosts to compare with the centroid velocity of the line. If the values are within 1-2 typical rotation curve velocities, we conclude that the hot halo material of the object caused the line. There are probably ways to distinguish between galaxies and groups, such as from ion ratios and column densities.

From the measured absorption line systems, we can extract a variety of physical properties about the hot gaseous component of galaxy halos and the intergalactic medium \citep{2015JATIS...1d5003B}. We can distinguish between density profiles of galaxies (i.e. NFW, $\beta = 1/2$, flat), determine the temperature from relative strengths of O VII and O VIII lines, and use temperature to determine the metallicity of the X-ray absorbing material. It is also important that we obtain estimates for the overall gas mass and dynamics of hot gaseous halos. These measurements might help to narrow the exact distribution of baryons in the universe, either verifying or ruling out existing models. Measurements of gas distribution from these absorption line studies will provide constraints on galaxy structure formation models as well. 

For an accurate measurement of these properties, it will be necessary to achieve greater sensitivity in X-ray spectroscopic technology. The \textit{Arcus} satellite is dedicated for such spectroscopy, 
with a proposed spectral resolution of $R\approx3000$ and an effective area of $A_{eff} \approx 300$ cm$^{2}$ \citep{2020SPIE11444E..2CS}, compared to \textit{Chandra} with $R\approx500$ and $A_{eff}\approx3.3$ cm$^{2}$ and \textit{XMM} with $R\approx420$ and $A_{eff}\approx45$ cm$^{2}$. This meets the sensitivity requirements to effectively determine gas dynamics and relative distances to absorbers.
If approved, \textit{Arcus} would begin observations late in the 2020-30 decade.
The \textit{Athena X-ray Observatory} \citep{barcons17} is under development with a planned launch date in the early 2030s.  For spectroscopy, it has a microcalorimeter (X-IFU) with a resolution of about 300 (2-2.5 eV; 1000 km s$^{-1}$) at low redshift for the O VII and O VIII lines.  Although this is an order of magnitude less than the \textit{Arcus} specifications, it has a collecting area of $ A_{eff} \approx 1.6$ m$^2$, so the detectability of lines is similar but at a loss of some velocity information. 
The \textit{Lynx} mission, if chosen, would fly probably in the early 2040s, and carries a higher resolution grating than \textit{Arcus}, with an expected resolution of about 5000 \citep{2019JATIS...5b1001G} and a proposed collecting area of $A_{eff} = 2$ m$^2$. 

Until the eventual launch of an advanced X-ray spectroscopic instrument, there is work that still needs to be done in preparation. It will be important to obtain the redshift and properties of potential galaxy hosts.  For the lower redshift systems, $z < 0.3$, this can be accomplished with 2-4 m class telescopes, but for more distant targets, 8 m class telescopes will be required.  The priority redshift range is $z < 1.3$, because at larger redshifts, the O VII resonance line is redshifted below the Galactic absorption cutoff. 

\section{Conclusion}

In this paper, we attempt to advance our strategy for detecting the missing baryons in hot X-ray absorbing material. By using NED data, and public optical DSS images, we produced separation vs. redshift plots, from which we identified the most likely absorbers in several AGN fields that X-ray observatories are likely to cover. From the results of our study, we determine critical steps for completing mission preparation and observing missing baryons:

• We need further optical and spectroscopic coverage of the fields we identified. This will give us enough accuracy to recognize the best AGN targets by the likelihood of observing absorption line systems in their spectra. 

• We can benefit from improving the precision in our technique for matching absorption lines to their respective hosts. 
One might gain improvements by comparing predictions from large scale structure simulations (e.g., \citealt{opp18b}, and testing it against the highest ionization UV lines, such as C IV and O VI.

The most likely absorption hosts identified here probably do not encompass every object that will cause absorption. There also may be a more detailed set of criteria, and when we have more complete optical/redshift coverage for objects this will be better determined. 

The future of X-ray absorption studies is arriving soon with high-resolution spectroscopy. Preparing  for this future is crucial for the success of these missions and for understanding the hot gaseous component of the universe.
\vspace{-0.3cm}
\section{Acknowledgements}

We thank the individuals who offered encouragement and insight, including Chris Miller, Sean Johnson, Mario Mateo, Laura Brennerman, Andrew Ptak, and Randall Smith.
We gratefully acknowledge support from NASA through the Astrophysics Data Analysis Program, awards NNX15AM93G and 80NSSC19K1013, and from the University of Michigan.
This research has made use of the NASA/IPAC Extragalactic Database (NED), which is operated by the Jet Propulsion Laboratory, California Institute of Technology,
under contract with NASA. 
This research has made use of the SIMBAD database and the Aladin sky atlas, which were produced and is maintained by the Centre de Données astronomiques de Strasbourg, France. 
We also acknowledge the use of several important sky surveys, which were produced with generous support from four US Federal agencies, more than a dozen organizations in eight countries, and two private foundations.  Specifically, these surveys include \textit{2MASS}, \textit{WISE}, \textit{SDSS}, \textit{Pan-STARRS}, the \textit{Dark Energy Survey}, and the \textit{VLT-ATLAS}. 
\appendix
\vspace{-0.5cm}
\section{Tables}
\begin{longrotatetable}
\begin{deluxetable*}{llllllll}
\tablenum{3}
\label{tbl3}
\tablecaption{Objects within 10$^\prime$ of 1ES 1028+511, sorted from least to greatest angular separation.}
\tablewidth{0pt}
\tablehead{ \colhead{Object} & \colhead{RA (h:m:s)}  & \colhead{DEC (d:m:s)}  & \colhead{Redshift}& \colhead{Type} & \colhead{App. mag} & \colhead{Sep. ($^\prime$)} & \colhead{Impact parameter (kpc)}
}
\startdata
WISEA J103121.30+505317.8 & 10:31:21.31 & +50:53:17.84  &                     &               & 0.533                  & 20           \\
WISEA J103113.72+505359.2 & 10:31:13.72 & +50:53:59.46  &                     &               & 0.853                  & 21.9         \\
WISEA J103112.90+505248.8 & 10:31:12.9  & +50:52:49.08  &                     &               & 1.179                  & 21.2         \\
WISEA J103112.41+505419.4 & 10:31:12.45 & +50:54:19.87  &                     &               & 1.205                  & 21.1         \\
SDSS J103113.45+505229.9  & 10:31:13.46 & +50:52:29.93  &                     &               & 1.357                  & 20.6         \\
WISEA J103108.50+505324.2 & 10:31:8.53  & +50:53:25.3   &                     &               & 1.585                  & 20.7         \\
2MASS J10312764+5052394   & 10:31:27.6  & +50:52:39     & 0.359698 & Spec          & 1.717                  & 21           & 795   \\
WISEA J103129.44+505310.5 & 10:31:29.42 & +50:53:10.39  &                     &               & 1.77                   & 21.9         \\
WISEA J103129.87+505337.3 & 10:31:29.86 & +50:53:37.79  &                     &               & 1.789                  & 19.9         \\
WISEA J103129.38+505406.9 & 10:31:29.41 & +50:54:6.91   &                     &               & 1.794                  & 20.7        
\enddata
\tablecomments{Table 3 is published in its entirety in the machine-readable format. We give a sample of 10 objects here.}
\end{deluxetable*}
\end{longrotatetable}
\begin{longrotatetable}
\begin{deluxetable*}{llllllll}
\tablenum{4}
\tablecaption{Objects within 10$^\prime$ of 1RXS J003334.6-192130, sorted from least to greatest angular separation.}
\tablewidth{0pt}
\tablehead{ \colhead{Object} & \colhead{RA (h:m:s)}  & \colhead{DEC (d:m:s)}  & \colhead{Redshift}& \colhead{Type} & \colhead{App. mag} & \colhead{Sep. ($^\prime$)} & \colhead{Impact parameter (kpc)}
}
\startdata
WISEA J003330.93-192108.8      & 00:33:30.94 & -19:21:9.04  &          &      & 0.87           & 19.29    \\
WISEA J003339.61-192138.0      & 00:33:39.65 & -19:21:37.91 &          &      & 1.287          & 19.8     \\
WISEA J003328.74-192053.4      & 00:33:28.73 & -19:20:53.92 &          &      & 1.448          & 19.5     \\
WISEA J003340.18-192238.1      & 00:33:40.23 & -19:22:38.42 &          &      & 1.79           & 19.85    \\
WISEA J003339.69-191952.6      & 00:33:39.73 & -19:19:53.44 &          &      & 2.115          & 19.5     \\
WISEA J003341.33-192011.9      & 00:33:41.34 & -19:20:11.54 &          &      & 2.166          & 19.51    \\
WISEA J003343.67-192104.4      & 00:33:43.72 & -19:21:4.18  &          &      & 2.298          & 19.83    \\
WISEA J003326.44-191951.1      & 00:33:26.45 & -19:19:51.1  &          &      & 2.499          & 19.1     \\
WISEA J003323.91-192008.9      & 00:33:23.88 & -19:20:9.46  &          &      & 2.807          & 19.73    \\
WISEA J003337.85-191847.2      & 00:33:37.96 & -19:18:47.16 &          &      & 2.908          & 19.9    
\enddata
\tablecomments{Table 4 is published in its entirety in the machine-readable format. We give a sample of 10 objects here. This field is outside the SDSS footprint, with a redshift close to the spectroscopic limit of $\approx0.5$. These factors make it a primary candidate for deeper spectroscopy and photometry.}
\label{tbl4}
\end{deluxetable*}
\end{longrotatetable}

\begin{longrotatetable}
\begin{deluxetable*}{llllllll}
\tablenum{5}
\label{tbl5}
\tablecaption{Objects within 10$^\prime$ of 1RXS J022716.6+0201541, sorted from least to greatest angular separation.}
\tablewidth{0pt}
\tablehead{ \colhead{Object} & \colhead{RA (h:m:s)}  & \colhead{DEC (d:m:s)}  & \colhead{Redshift}& \colhead{Type} & \colhead{App. mag} & \colhead{Sep. ($^\prime$)} & \colhead{Impact parameter (kpc)}
}
\startdata
WISEA J022712.68+020111.6      & 02:27:12.62 & 02:01:11.28 & 0.188566 & Spec & 19.45    & 1.286           \\
WISEA J022716.94+015948.9      & 02:27:16.93 & 01:59:48.98 &                     &      & 18.5     & 2.193           \\
WISEA J022707.38+020019.7      & 02:27:7.35  & 02:00:19.69 & 0.188994 & Spec & 19.15    & 2.853           \\
WISEA J022705.29+020006.7      & 02:27:5.22  & 02:00:6.08  &                     &      & 20.37    & 3.419           \\
WISEA J022720.93+020614.6      & 02:27:20.94 & 02:06:14.51 & 0.387021 & Spec & 16.907   & 4.371           \\
APMUKS(BJ) B022432.88+014436.3 & 02:27:7.92  & 01:58:2.39  &                     &      & 20.49    & 4.52            \\
UGC 01923                      & 02:27:25.55 & 02:06:18.11 & 0.009593 & Spec & 17       & 4.843           & 57.4  \\
WISEA J022710.50+015645.5      & 02:27:10.45 & 01:56:45.46 &                     &      & 20.4     & 5.469           \\
WISEA J022652.73+015937.0      & 02:26:52.73 & 01:59:37.1  & 0.362079 & Spec &          & 6.421           \\
WISEA J022739.31+015825.3      & 02:27:39.28 & 01:58:25.82 &                     &      & 19.32    & 6.705          
\enddata
\tablecomments{Table 5 is published in its entirety in the machine-readable format. We give a sample of 10 objects here. This source lies outside the SDSS footprint, and the surrounding objects are not well documented. In addition to the field described in Table 4, this is another prime candidate for deeper imaging.}
\end{deluxetable*}
\end{longrotatetable}

\begin{longrotatetable}
\begin{deluxetable*}{llllllll}
\tablenum{6}
\label{tbl6}
\tablecaption{Objects within 10$^\prime$ of 1RXS J110337.7-232931, sorted from least to greatest angular separation.}
\tablewidth{0pt}
\tablehead{ \colhead{Object} & \colhead{RA (h:m:s)}  & \colhead{DEC (d:m:s)}  & \colhead{Redshift}& \colhead{Type} & \colhead{App. mag} & \colhead{Sep. ($^\prime$)} & \colhead{Impact parameter (kpc)}
}
\startdata
H   1101-232G1               & 11:03:32.5 & -23:30:44   & 0.187    & Spec & 18.1     & 1.675            \\
GALEXASC J110330.49-233059.6 & 11:03:30.5 & -23:30:58   & 0.14     & Spec & 21.34    & 2.168            \\
WISEA J110335.26-233214.2    & 11:03:35.2 & -23:32:14   &          &      &          & 2.774            \\
WISEA J110331.98-232617.9    & 11:03:32.0 & -23:26:18   &          &      &          & 3.471            \\
WISEA J110322.59-233521.1    & 11:03:22.6 & -23:35:21   &          &      &          & 6.776            \\
WISEA J110330.22-233635.2    & 11:03:30.2 & -23:36:35   &          &      &          & 7.272              
\enddata
\tablecomments{Table 6 is published in its entirety in the machine-readable format. We give a sample of 10 objects here. This source lies outside the SDSS footprint, and the surrounding objects are not well documented. In addition to the field described in Table 4, this is another prime candidate for deeper imaging.}
\end{deluxetable*}
\end{longrotatetable}

\begin{longrotatetable}
\begin{deluxetable*}{llllllll}
\tablenum{7}
\label{tbl7}
\tablecaption{Objects within 10$^\prime$ of 1RXS J111706.3+201410, sorted from least to greatest angular separation.}
\tablewidth{0pt}
\tablehead{ \colhead{Object} & \colhead{RA (h:m:s)}  & \colhead{DEC (d:m:s)}  & \colhead{Redshift}& \colhead{Type} & \colhead{App. mag} & \colhead{Sep. ($^\prime$)} & \colhead{Impact parameter (kpc)}
}
\startdata
2MASS J11170682+2014039   & 11:17:6.83  & 20:14:3.59  &                     &      & 19.9     & 0.148            \\
SDSS J111705.91+201417.5  & 11:17:5.92  & 20:14:17.52 &                     &      & 20.4     & 0.187            \\
WISEA J111705.49+201412.9 & 11:17:5.48  & 20:14:11.58 &                     &      & 19.1     & 0.195            \\
WISEA J111708.46+201401.7 & 11:17:8.48  & 20:14:1.68  &                     &      & 20.7     & 0.529            \\
SDSS J111705.25+201440.8  & 11:17:5.26  & 20:14:40.85 &                     &      & 20.3     & 0.605            \\
WISEA J111704.66+201442.8 & 11:17:4.63  & 20:14:42.83 &                     &      & 19.7     & 0.703            \\
WISEA J111710.11+201401.1 & 11:17:10.11 & 20:14:0.96  &                     &      & 20.8     & 0.909            \\
WISEA J111709.83+201337.8 & 11:17:9.86  & 20:13:37.78 &                     &      & 21       & 0.977            \\
WISEA J111710.91+201355.3 & 11:17:10.91 & 20:13:55.02 &                     &      & 19.5     & 1.109            \\
SDSS J111711.15+201359.2  & 11:17:11.16 & 20:13:59.3  &                     &      & 20.5     & 1.157            
\enddata
\tablecomments{Table 7 is published in its entirety in the machine-readable format. We give a sample of 10 objects here. The objects in this field have mostly photo-z redshift values, so spectroscopic followup will be necessary.}
\end{deluxetable*}
\end{longrotatetable}

\begin{longrotatetable}
\begin{deluxetable*}{llllllll}
\tablenum{8}
\label{tbl8}
\tablecaption{Objects within 15$^\prime$ of 1RXS J151747.3+652522, sorted from least to greatest angular separation.}
\tablewidth{0pt}
\tablehead{ \colhead{Object} & \colhead{RA (h:m:s)}  & \colhead{DEC (d:m:s)}  & \colhead{Redshift}& \colhead{Type} & \colhead{App. mag} & \colhead{Sep. ($^\prime$)} & \colhead{Impact parameter (kpc)}
}
\startdata
WISEA   J151748.75+652456.8 & 15:17:48.93 & 65:24:58.0  &                 &      & 20.40      & 0.454  \\
WISEA J151754.35+652342.2   & 15:17:54.37 & 65:23:42.22 &                 &      & 20.80      & 1.835  \\
SDSS J151729.52+652543.6    & 15:17:29.52 & 65:25:43.68 &                 &      & 20.40      & 1.908  \\
2MASX J15175301+6523256     & 15:17:53.07 & 65:23:26.52 &0.061792
                 &Phot      & 16.47      & 2.038  \\
WISEA J151759.90+652348.8   & 15:17:59.96 & 65:23:48.84 &                 &      & 20.30      & 2.041  \\
WISEA J151810.21+652529.7   & 15:18:10.2  & 65:25:29.68 &                 &      & 20.30      & 2.352  \\
WISEA J151721.63+652409.8   & 15:17:21.8  & 65:24:10.94 &                 &      & 20.90      & 2.946  \\
SDSS J151813.00+652658.9    & 15:18:13.01 & 65:26:58.96 &                 &      & 20.10      & 3.079  \\
WISEA J151812.77+652706.4   & 15:18:12.72 & 65:27:6.41  &                 &      & 20.40      & 3.12   
\enddata
\tablecomments{Table 8 is published in its entirety in the machine-readable format. We give a sample of 10 objects here. The only objects with available redshift values lie outside a 10$^\prime$ radius of the source. Spectroscopic followup with objects in this field within 10$^\prime$ are necessary to complete this field.}
\end{deluxetable*}
\end{longrotatetable}

\begin{longrotatetable}
\begin{deluxetable*}{llllllll}
\tablenum{9}
\label{tbl9}
\tablecaption{Objects within 10$^\prime$ of 1RXS J153501.1+532042, sorted from least to greatest angular separation.}
\tablewidth{0pt}
\tablehead{ \colhead{Object} & \colhead{RA (h:m:s)}  & \colhead{DEC (d:m:s)}  & \colhead{Redshift}& \colhead{Type} & \colhead{App. mag} & \colhead{Sep. ($^\prime$)} & \colhead{Impact parameter (kpc)}
}
\startdata
WISEA   J153500.26+532000.7 & 15:35:00.2  & +53:20:00   & 0.0293361    & Spec & 16.5 & 0.631      & 24.895\\
WISEA J153506.70+532029.2   & 15:35:06.7  & +53:20:29   & 0.466693 & Spec & 22   & 0.893      & 537\\
WISEA J153455.37+531952.6   & 15:34:55.39 & 53:19:52.61 &                     &      & 19.8 & 1.1        \\
WISEA J153509.26+531948.7   & 15:35:9.27  & 53:19:48.9  &                     &      & 20.3 & 1.5        \\
WISEA J153504.05+531825.2   & 15:35:4.06  & 53:18:25.16 &                     &      & 20.2 & 2.256      \\
WISEA J153515.89+532051.4   & 15:35:15.91 & 53:20:51.54 &                     &      & 20.7 & 2.267      \\
SDSS J153447.51+531914.6    & 15:34:47.51 & 53:19:14.7  &                     &      & 19.6 & 2.415      \\
WISEA J153458.46+532305.5   & 15:34:58.44 & 53:23:5.32  &                     &      & 20.7 & 2.491      \\
WISEA J153456.60+531813.1   & 15:34:56.58 & 53:18:12.46 &                     &      & 19.9 & 2.495      \\
SDSS J153510.47+531810.8    & 15:35:10.48 & 53:18:10.84 &                     &      & 20.7 & 2.837      
\enddata
\tablecomments{Table 9 is published in its entirety in the machine-readable format. We give a sample of 10 objects here.}
\end{deluxetable*}
\end{longrotatetable}

\begin{longrotatetable}
\begin{deluxetable*}{llllllll}
\tablenum{10}
\label{tbl10}
\tablecaption{Objects within 10$^\prime$ of 3C 279, sorted from least to greatest angular separation.}
\tablewidth{0pt}
\tablehead{ \colhead{Object} & \colhead{RA (h:m:s)}  & \colhead{DEC (d:m:s)}  & \colhead{Redshift}& \colhead{Type} & \colhead{App. mag} & \colhead{Sep. ($^\prime$)} & \colhead{Impact parameter (kpc)}
}
\startdata
WISEA J125611.81-054535.6      & 12:56:11.86 & -05:45:35.75 &                     &      & 19.77    & 1.771            \\
WISEA J125609.77-054458.7      & 12:56:9.59  & -05:44:56.98 &                     &      & 19.55    & 2.441            \\
LCRS B125323.5-052906          & 12:55:58.73 & -05:45:21.56 &                     &      & 17.94    & 3.684            \\
WISEA J125607.81-055115.7      & 12:56:7.85  & -05:51:15.73 &                     &      & 19.77    & 3.99             \\
LCRS B125328.2-053447          & 12:56:3.6   & -05:51:1.19  &                     &      & 17.99    & 4.117            \\
WISEA J125617.41-055117.2      & 12:56:17.42 & -05:51:17.46 &                     &      & 19.2     & 4.229            \\
LCRS B125328.0-052716          & 12:56:3.3   & -05:43:30.5  &                     &      & 18.44    & 4.319            \\
WISEA J125611.51-055206.2      & 12:56:11.52 & -05:52:6.35  &                     &      & 20.36    & 4.748            \\
WISEA J125557.93-054347.6      & 12:55:57.94 & -05:43:48.11 &                     &      & 20.09    & 4.845            \\
LCRS B125355.2-053144          & 12:56:30.5  & -05 47 58    & 0.106774 & Spec & 15.242   & 4.847            
\enddata
\tablecomments{Table 10 is published in its entirety in the machine-readable format. We give a sample of 10 objects here. This field lies outside the SDSS footprint has poor spectroscopic coverage. Deeper imaging is needed to complete the information about this field}
\end{deluxetable*}
\end{longrotatetable}

\begin{longrotatetable}
\begin{deluxetable*}{llllllll}
\tablenum{11}
\label{tbl11}
\tablecaption{Objects within 10$^\prime$ of 3C 454.3, sorted from least to greatest angular separation.}
\tablewidth{0pt}
\tablehead{ \colhead{Object} & \colhead{RA (h:m:s)}  & \colhead{DEC (d:m:s)}  & \colhead{Redshift}& \colhead{Type} & \colhead{App. mag} & \colhead{Sep. ($^\prime$)} & \colhead{Impact parameter (kpc)}
}
\startdata
3C 454.3:{[}KSS94{]} 28            & 22:53:57.8  & 16:08:39.01 &                         &      & 23.1      & 0.243               \\
WISEA J225355.74+160850.7          & 22:53:55.7  & 16:08:49.99 &                         &      & 22.2      & 0.495               \\
{[}CLW2001{]} 2251+1552 +0322-0003 & 22:53:59.94 & 16:08:53.3  & 0.3529 & Spec &           & 0.525               \\
RX J2253.9+1608:{[}ZEH2003{]} 04   & 22:54:0.0   & 16:08:30.98 &                         &      &           & 0.659               \\
SSTSL2 J225400.41+160906.4         & 22:54:0.5   & 16:09:6.01  & 0.3526 & Spec & 21.6      & 0.693               \\
3C 454.3:{[}KSS94{]} 17            & 22:53:55.4  & 16:09:25.99 &                         &      & 22        & 0.781               \\
GALEXMSC J225356.94+160944.6       & 22:53:57.1  & 16:09:42.98 &                         &      & 21.7      & 0.839               \\
3C 454.3:{[}KSS94{]} 05            & 22:53:58.2  & 16:09:43.99 &                         &      & 22.8      & 0.848               \\
SSTSL2 J225400.57+160925.5         & 22:54:0.6   & 16:09:24.98 &                         &      & 22.1      & 0.862               \\
WISEA J225355.89+160938.9          & 22:53:55.9  & 16:09:38.02 &                         &      & 21        & 0.863               
\enddata
\tablecomments{Table 11 is published in its entirety in the machine-readable format. We give a sample of 10 objects here.}
\end{deluxetable*}
\end{longrotatetable}

\begin{longrotatetable}
\begin{deluxetable*}{llllllll}
\tablenum{12}
\label{tbl12}
\tablecaption{Objects within 10$^\prime$ of H2356-309, sorted from least to greatest angular separation.}
\tablewidth{0pt}
\tablehead{ \colhead{Object} & \colhead{RA (h:m:s)}  & \colhead{DEC (d:m:s)}  & \colhead{Redshift}& \colhead{Type} & \colhead{App. mag} & \colhead{Sep. ($^\prime$)} & \colhead{Impact parameter (kpc)}
}
\startdata
WISEA J235910.71-303720.6      & 23:59:10.73 & -30:37:21.61 & 0.0627  & Spec & 18.7                 & 0.684               \\
MRSS 409-104402                & 23:59:14.62 & -30:37:24.67 &                     &      & 20                   & 1.467               \\
WISEA J235913.66-303645.8      & 23:59:13.67 & -30:36:45.97 &                     &      & 20.47                & 1.537               \\
WISEA J235911.47-303908.4      & 23:59:11.37 & -30:39:7.09  &                     &      & 18.8                 & 1.622               \\
WISEA J235903.56-303906.3      & 23:59:3.71  & -30:39:8.28  &                     &      & 19.43                & 1.718               \\
MRSS 409-105369                & 23:59:6.35  & -30:35:49.7  &                     &      & 18.6                 & 1.878               \\
MRSS 409-104126                & 23:59:17.26 & -30:37:43.32 &                     &      & 20                   & 2.011               \\
WISEA J235901.67-303926.5      & 23:59:1.67  & -30:39:25.99 &                     &      & 19.8                 & 2.21                \\
WISEA J235917.55-303651.0      & 23:59:17.49 & -30:36:50.8  &                     &      & 19.4                 & 2.222               \\
WISEA J235917.89-303833.5      & 23:59:17.94 & -30:38:33.86 &                     &      & 18.3                 & 2.332               
\enddata
\tablecomments{Table 12 is published in its entirety in the machine-readable format. We give a sample of 10 objects here.}
\end{deluxetable*}
\end{longrotatetable}

\begin{longrotatetable}
\begin{deluxetable*}{llllllll}
\tablenum{13}
\label{tbl13}
\tablecaption{Objects within 10$^\prime$ of S5 0836+71, sorted from least to greatest angular separation.}
\tablewidth{0pt}
\tablehead{ \colhead{Object} & \colhead{RA (h:m:s)}  & \colhead{DEC (d:m:s)}  & \colhead{Redshift}& \colhead{Type} & \colhead{App. mag} & \colhead{Sep. ($^\prime$)} & \colhead{Impact parameter (kpc)}
}
\startdata
WISEA J084140.99+705210.4         & 08:41:40.84 & 70:52:9.98  &                     &      & 18.86 & 2.044               \\
WISEA J084047.07+705544.2         & 08:40:46.95 & 70:55:45.01 &                     &      & 19.24 & 3.681               \\
GALEXASC J084150.37+705009.3      & 08:41:50.4  & +70:50:09   & 1.242               & Phot & 18.3  & 4.147               \\
WISEA J084213.74+705146.0         & 08:42:13.56 & 70:51:46.01 &                     &      & 18.75 & 4.47                \\
WISEA J084134.24+704906.8         & 08:41:34.01 & 70:49:7.0   &                     &      & 19.32 & 4.654               \\
WISEA J084216.83+705148.6         & 08:42:16.64 & 70:51:47.99 &                     &      & 19.31 & 4.685               \\
WISEA J084030.58+705117.9         & 08:40:30.4  & 70:51:19.01 &                     &      & 18.3  & 5.023               \\
{[}VMF98{]} 057:{[}BBS2002{]} 051 & 08:40:52.3  & +70:49:17   & 0.2418              & Spec & 19.5  & 5.138               \\
2MASS J08415508+7048592           & 08:41:54.7  & +70:48:59   & 0.2387              & Spec & 19.13 & 5.337               \\
WISEA J084124.44+704820.1         & 08:41:24.21 & 70:48:20.02 &                     &      & 19.02 & 5.37                
\enddata
\tablecomments{Table 13 is published in its entirety in the machine-readable format. We give a sample of 10 objects here.}
\end{deluxetable*}
\end{longrotatetable}

\begin{longrotatetable}
\begin{deluxetable*}{llllllll}
\tablenum{14}
\label{tbl14}
\tablecaption{Objects within 10$^\prime$ of Ton 116, sorted from least to greatest angular separation.}
\tablewidth{0pt}
\tablehead{ \colhead{Object} & \colhead{RA (h:m:s)}  & \colhead{DEC (d:m:s)}  & \colhead{Redshift}& \colhead{Type} & \colhead{App. mag} & \colhead{Sep. ($^\prime$)} & \colhead{Impact parameter (kpc)}
}
\startdata
WISEA J124314.74+362752.6       & 12:43:14.7  & 36:27:53.35 &                     &      & 20.6      & 0.424      \\
RX J1243.2+3627:{[}BEV98{]} 002 & 12:43:14.8  & 36:27:51.01 &                     &      & 18.7      & 0.431      \\
WISEA J124316.96+362713.6       & 12:43:16.97 & 36:27:13.79 & 0.485129 & Spec & 21.3      & 0.989      & 618\\
WISEA J124305.68+362654.3       & 12:43:5.7   & 36:26:54.2  &                     &      & 20.8      & 1.641      \\
WISEA J124321.34+362810.8       & 12:43:21.33 & 36:28:11.03 &                     &      & 20.8      & 1.787      \\
SDSS J124312.06+362544.1        & 12:43:12.06 & 36:25:44.15 &                     &      & 20.9      & 2.002      \\
WISEA J124321.71+362839.8       & 12:43:21.72 & 36:28:39.9  &                     &      & 20.8      & 2.033      \\
WISEA J124304.77+362617.8       & 12:43:4.81  & 36:26:17.74 &                     &      & 20.4      & 2.147      \\
WISEA J124315.69+362952.9       & 12:43:15.72 & 36:29:53.27 &                     &      & 19.6      & 2.236      \\
WISEA J124323.65+362635.4       & 12:43:23.6  & 36:26:35.23 &                     &      & 20.9      & 2.467                                 
\enddata
\tablecomments{Table 13 is published in its entirety in the machine-readable format. We give a sample of 10 objects here.}
\end{deluxetable*}
\end{longrotatetable}

\begin{longrotatetable}
\begin{deluxetable*}{llllllll}
\tablenum{15}
\label{tbl15}
\tablecaption{Objects within 10$^\prime$ of 1RXS J122121.7+301041, sorted from least to greatest angular separation.}
\tablewidth{0pt}
\tablehead{ \colhead{Object} & \colhead{RA (h:m:s)}  & \colhead{DEC (d:m:s)}  & \colhead{Redshift}& \colhead{Type} & \colhead{App. mag} & \colhead{Sep. ($^\prime$)} & \colhead{Impact parameter (kpc)}
}
\startdata
WISEA J122122.35+301021.1 & 12:21:22.34 & 30:10:20.78 &                     &      & 20.2                 & 0.285      \\
WISEA J122125.49+300932.7 & 12:21:25.49 & 30:09:32.87 &                     &      & 19.6                 & 1.317      \\
SDSS J122129.52+301103.6  & 12:21:29.52 & 30:11:3.66  &                     &      & 20.9                 & 1.697      \\
WISEA J122123.38+300853.6 & 12:21:23.39 & 30:08:53.7  &                     &      & 19.2                 & 1.751      \\
SDSS J122115.97+300924.9  & 12:21:15.97 & 30:09:24.98 &                     &      & 20.7                 & 1.763      \\
WISEA J122114.21+300925.5 & 12:21:14.18 & 30:09:25.96 &                     &      & 20.7                 & 2.054      \\
WISEA J122132.81+301145.6 & 12:21:32.78 & 30:11:44.77 &                     &      & 20.1                 & 2.6        \\
WISEA J122134.81+301123.1 & 12:21:34.84 & 30:11:23.39 & 0.060387 & Spec & 18.3                 & 2.892      & 230\\
WISEA J122111.10+301218.6 & 12:21:11.09 & 30:12:18.79 &                     &      & 20.9                 & 2.893      \\
WISEA J122134.44+301149.1 & 12:21:34.48 & 30:11:49.13 &                     &      & 18.49                & 2.963                                      
\enddata
\tablecomments{Table 14 is published in its entirety in the machine-readable format. We give a sample of 10 objects here.}
\end{deluxetable*}
\end{longrotatetable}

\begin{longrotatetable}
\begin{deluxetable*}{llllllll}
\tablenum{16}
\label{tbl16}
\tablecaption{Objects within 10$^\prime$ of 1RXS J142239.1+580159, sorted from least to greatest angular separation.}
\tablewidth{0pt}
\tablehead{ \colhead{Object} & \colhead{RA (h:m:s)}  & \colhead{DEC (d:m:s)}  & \colhead{Redshift}& \colhead{Type} & \colhead{App. mag} & \colhead{Sep. ($^\prime$)} & \colhead{Impact parameter (kpc)}
}
\startdata
WISEA J142325.43+575923.7 & 14:23:25.43 & 57:59:23.5  & 0.04411  & Spec & 17.1                 & 6.663      & 388\\
SDSS J142237.38+580036.1  & 14:22:37.38 & 58:00:36.14 &                     &      & 20                   & 1.338      \\
WISEA J142245.79+580026.4 & 14:22:45.78 & 58:00:26.28 &                     &      & 20.8                 & 1.744      \\
WISEA J142222.64+580122.1 & 14:22:22.7  & 58:01:21.18 &                     &      & 20.5                 & 2.218      \\
WISEA J142251.82+580324.2 & 14:22:51.85 & 58:03:24.19 &                     &      & 19.1                 & 2.263      \\
WISEA J142245.79+580408.5 & 14:22:45.81 & 58:04:8.98  &                     &      & 20.6                 & 2.405      \\
SDSS J142224.44+575959.6  & 14:22:24.45 & 57:59:59.64 &                     &      & 20.9                 & 2.718      \\
WISEA J142300.70+580206.3 & 14:23:0.72  & 58:02:5.82  &                     &      & 20.4                 & 2.894      \\
WISEA J142232.79+575905.3 & 14:22:32.85 & 57:59:5.57  &                     &      & 20.3                 & 2.943      \\
WISEA J142216.63+580211.3 & 14:22:16.65 & 58:02:11.83 &                     &      & 20.2                 & 2.957                                         
\enddata
\tablecomments{Table 15 is published in its entirety in the machine-readable format. We give a sample of 10 objects here.}
\end{deluxetable*}
\end{longrotatetable}

\begin{longrotatetable}
\begin{deluxetable*}{llllllll}
\tablenum{17}
\label{tbl17}
\tablecaption{Objects within 10$^\prime$ of 1RXS J150759.8+041511, sorted from least to greatest angular separation.}
\tablewidth{0pt}
\tablehead{ \colhead{Object} & \colhead{RA (h:m:s)}  & \colhead{DEC (d:m:s)}  & \colhead{Redshift}& \colhead{Type} & \colhead{App. mag} & \colhead{Sep. ($^\prime$)} & \colhead{Impact parameter (kpc)}
}
\startdata
WISEA J150800.69+041332.0 & 15:08:0.7   & 04:13:32.02 & 0.140261 & Spec & 18.5                 & 1.684      & 307\\
WISEA J150804.48+041420.7 & 15:08:4.51  & 04:14:21.05 &                     &      & 20                   & 1.462      \\
SDSS J150755.09+041414.1  & 15:07:55.09 & 04:14:14.14 &                     &      & 19                   & 1.506      \\
SDSS J150754.03+041438.8  & 15:07:54.04 & 04:14:38.9  &                     &      & 20.3                 & 1.523      \\
SDSS J150804.76+041413.7  & 15:08:4.77  & 04:14:13.74 &                     &      & 20.7                 & 1.587      \\
WISEA J150754.07+041602.8 & 15:07:54.07 & 04:16:2.78  &                     &      & 20.8                 & 1.646      \\
WISEA J150801.68+041328.6 & 15:08:1.69  & 04:13:28.7  &                     &      & 20.8                 & 1.789      \\
WISEA J150753.76+041411.6 & 15:07:53.8  & 04:14:9.82  &                     &      & 20.5                 & 1.805      \\
WISEA J150805.58+041618.6 & 15:08:5.58  & 04:16:19.31 &                     &      & 19.5                 & 1.839      \\
WISEA J150756.46+041701.4 & 15:07:56.45 & 04:17:1.64  & 0.775    & Spec & 20.047               & 2.002      & 1010
\enddata
\tablecomments{Table 17 is published in its entirety in the machine-readable format. We give a sample of 10 objects here.}
\end{deluxetable*}
\end{longrotatetable}

\bibliography{absorptionhosts}{}
\bibliographystyle{aasjournal}
\end{document}